\documentclass[preprint2,longabstract]{aastex}

\shorttitle{Filaments in far infrared and sub-mm images}
\shortauthors{Schisano~E., Rygl~K.~L.~J., Molinari~S. et al.}

\usepackage{natbib}
\usepackage{graphicx}
\usepackage{txfonts}
\bibpunct{(}{)}{;}{a}{}{,} 

\newcommand{\lone}{$\lambda_{1}$}
\newcommand{\ltwo}{$\lambda_{2}$}
\newcommand{\cmtwo}{cm$^{-2}$}

\newcommand{\um}{$\mu$m}                                 
\newcommand{\msun}{$M_{\odot}$}

\newcommand{\msunpc}{$M_{\odot}$\,pc$^{-1}$}

\newcommand{\lsim}{\;\lower.6ex\hbox{$\sim$}\kern-7.75pt\raise.65ex\hbox{$<$}\;}
\newcommand{\gsim}{\;\lower.6ex\hbox{$\sim$}\kern-7.75pt\raise.65ex\hbox{$>$}\;}
\newcommand{\gl}{\;\lower.6ex\hbox{$<$}\kern-7.75pt\raise.65ex\hbox {$>$}\;}
\newcommand{\eg}    {e.\,g.,}
\newcommand{\ie}     {i.\,e.,}
\newcommand{\mlcrit}	{$M_{\mathrm{line,crit}}$}
\newcommand{\mline}	{$M_{\mathrm{line}}$}

\begin{document}
   \title{The identification of filaments on far infrared and submillimiter images. \\
	Morphology, physical conditions and relation with star formation of filamentary structure }

   \author{E.~Schisano\altaffilmark{1}}
   	\affil{Infrared Processing and Analysis Center, California Institute of Technology, Pasadena, CA, 91125, USA.}
	\email{eugenio@ipac.caltech.edu}
 	\altaffiltext{1}{Istituto di Astrofisica e Planetologia Spaziali, INAF-IAPS, Via Fosso del Cavaliere 100, 00133, Roma, Italy.}

  \author{K.~L.~J. Rygl\altaffilmark{1}} 
  	\affil{European Space Research and Technology Centre (ESA-ESTEC), Keplerlaan 1, P.O. Box 299, 2200 AG Noordwijk, The Netherlands}
 
  \author{S.~Molinari and G.~Busquet and D.~Elia and M.~Pestalozzi}
  	\affil{Istituto di Astrofisica e Planetologia Spaziali, INAF-IAPS, Via Fosso del Cavaliere 100, 00133, Roma, Italy.}
	
  \author{D.~Polychroni\altaffilmark{1}}
	\affil{University of Athens, Departement of Astrophysics, Astronomy and Mechanics, Faculty of Physics, Panepistimiopolis, 15784 Zografos, Athens, Greece.} 

  \author{N.~Billot}
          \affil{Instituto de RadioAstronom\'{i}a Milim\'{e}trica Avenida Divina Pastora, 7, N\'{u}cleo Central E 18012 Granada, Spain}
  
  \author{S.~Carey  and R.~Paladini}
	\affil{Infrared Processing and Analysis Center, California Institute of Technology, Pasadena, CA, 91125, USA.}

  \author{ A.~Noriega-Crespo}
	\affil{Space Telescope Science Institute, Baltimore, 21218, USA}  

  \author{T.~J.~T. Moore}
	\affil{Astrophysics Research Institute, Liverpool John Moores University, 146 Brownlow Hill, Liverpool, L3 5RF UK. }

  \author{R.~Plume}
 	\affil{Department of Physics and Astronomy and the Institute for Space Imaging Sciences, University of Calgary, Calgary, AB T2N IN4, Canada.}

  \author{S.~C.~O. Glover}
	\affil{Zentr\"{u}m f\"{u}r Astronomie, Institut f\"{u}r Theoretische Astrophysik, Universit\"{a}t Heidelberg, Albert-Ueberle-Str. 2, D-69120 Heidelberg, Germany.}  

  \and
  \author{E.~V\'{a}zquez-Semadeni}
      	\affil{Centro de Radioastronom\'{i}a y Astrof\'{i}sica (CRyA), Universidad Nacional Aut\'{o}noma de M\'{e}xico, CP 58190 Morelia, Michoac\'{a}n, Mexico.}

  \begin{abstract}
   Observations of molecular clouds reveal a complex structure, with
   gas and dust often arranged in filamentary rather than spherical geometries. 
   The associations of pre- and proto- stellar cores with
   the filaments suggest a direct link with the process of star formation.
   Any study of the properties of such filaments requires a representative samples
   from different enviroments and so an unbiased detection method.
   We developed such an approach using the Hessian matrix of a surface-brightness 
   distribution to identify filaments and determine their physical and morphological properties. 
   After testing the method on simulated, but realistic filaments, we apply the algorithms to column-density maps 
   computed from {\it Herschel} observations of the Galactic Plane obtained by the Hi-GAL project. 
   We identified $\sim$ 500 filaments, in the longitude range of $l$=216.5$^{\rm o}$ to $l$=225.5$^{\rm o}$, 
   with lengths from $\sim$1\,pc up to $\sim$30\,pc and widths between 0.1\,pc and 2.5\,pc. 
   Average column densities are between 10$^{20}$\,\cmtwo and 10$^{22}$\,\cmtwo. 
   Filaments include the majority of dense material with $N_{\rm H_{2}}$$>$6$\times$10$^{21}$\,\cmtwo.
   We find that the pre- and proto-stellar compact sources already identified in the same region are mostly associated
   with filaments. However, surface densities in excess of the expected critical values for  high-mass star formation are 
   only found on the filaments, indicating that these structures are necessary to channel material into the clumps. 
   Furthermore, we analyze the gravitational stability of filaments and discuss their relationship with star formation.

   \end{abstract}
  

   \maketitle

\section{Introduction}

Molecular clouds are the birthplaces of stars. Observations at different 
wavelengths and using different molecular tracers of the relatively best studied and nearby star-forming regions suggest complex morphologies, with the dust and gas arranged mostly along elongated, almost one-dimensional, filamentary structures \citep[\eg][]{Hartmann2002,Hatchell2005,Myers2009}. The {\em Herschel} Space Observatory \citep{pilbratt10}, thanks to its superior spatial resolution and sensitivity in the far-infrared, is now showing that the filamentary organization of the dense interstellar material is much more pervasive than was initially thought. From sub-parsec scales in nearby star formation regions \citep{Andre2010} to tens-of-parsecs scale along spiral arms \citep{molinari2010}, filaments appear to be key-structures required to build the densities necessary for star formation. The abundance of compact star-forming seeds along these structures \citep[see][]{elia10,Hennings2010}, from pre-stellar to protostellar young condensations, indicates that it is in filaments where the initial conditions 
for star formation may be set.

Despite the ubiquity of filaments in star forming regions, it is still unclear how they form and 
what their real relationship is with the mechanisms of star formation. Recent theoretical modeling of molecular cloud formation tends to produce filamentary structures formed by different mechanisms, like decaying supersonic turbulence \citep{Padoan2007}, cooling in the post-shock regions of large-scale colliding flows \citep{Heitch2008, Vazquez2011}, or global gravitational instabilities \citep{Hartmann2007}. While these predictions seem in qualitative agreement with observed morphologies, a detailed quantitative comparison has yet to be done. Significant advances are now possible with the availability of complete panoramic surveys of both the Galactic Plane, the Hi-GAL project \citep{molinari2010}, and of nearby star-forming regions, like the Gould Belt project \citep{Andre2010}, carried out with the {\em Herschel} satellite that, in principle, allow an unbiased characterization of filaments over a wide range of spatial scales and physical conditions. 

Visual selection methods used in the recent past to identify the most obvious structures that appear elongated and the subsequent manual analyses to selected
portions of these filaments \citep[\eg][]{Hartmann2002,Hatchell2005,busquet2013} become impratical when applied to large data sets. For example, the Hi-GAL project, data from which has been used for this work, mapped with {\it Herschel} the entire Galactic Plane at 70, 160, 250, 350, and 500 $\mu$m covering a total area of $\sim$720 square degrees \citep{molinari2010a}. Filaments are found everywhere in Hi-GAL maps. Therefore, the quantitative and qualitative order-of-magnitude improvement in available datasets brought by {\em Herschel}, requires a change of perspectives when it comes to analysis methodologies.
The problem of identifying specific patterns in images has  already been faced in other scientific fields, in particular in computer engineering \citep[see for example][]{GW2002}, using \textit{ad hoc} convolution  with optimal filtering (like the Canny detector; \citealt{Canny1986}) or studies of the local properties (topology) of the images (Hessian matrix studies, Skeleton, Morse Theory, Shapefinders; \citealt{Sheth2003}). More related to astrophysics is the issue of determining the filamentary pattern from cosmological N-body simulations of the dark matter distributions (the cosmic web) or from the observed large-scale distribution of galaxies. To accomplish these goals,
different approaches have been developed \cite[see][for a review]{AragonCalvo2007} reaching different degrees of complexity. 
 
More recently, \cite{Sousbie11} has presented a specific formalism (DisPerSE) based on the discrete Morse theory, which is able to recognize salient features of the large scale cosmic web. The corresponding software has been already applied successfully to column density maps computed from the far-infrared/sub-millimeter data \citep[\eg][]{Arzoumanian2011,Hill2011,Peretto2012}. Nevertheless, the key issue to be addressed when identifying particular patterns is a definition of the feature to be identified. Given a precise definition for the desired pattern, it is possible to define the best method to highlight the defined structures. As an example in the skeleton approach, as well as in DisPerSE, a filament is defined as the one-dimensional segment given by the central denser region of extended elliptical structure. For this reason, the skeleton is determined by choosing among all the paths that connect the saddle points of the density (intensity) field to the local maxima, the one that, point by point, shows the smallest variation in the gradient \citep{Novikov2006, Sousbie11}.
Such a definition allows the correct tracing of the ridge of the filaments.

In this paper, we consider filaments not  as a one-dimensional structures for which we simply trace the main ridge, or spine, but as an extended 2-dimensional feature that covers a portion of the map.  
Our aim is then to identify on the map the regions that belong to the filamentary structure in order to derive its morphological and physical properties. 
To this end, we start by defining a filament as {\it an elongated region with a relatively higher brightness contrast with respect to its surrounding}, formalizing the intuitive idea of what a filament looks like based on what the eye sees on a map. Hence, instead of an approach involving the local extrema, we prefer to focus on a differential method, specifically the investigation of the eigenvalues of the Hessian matrix of the intensity (density) field, directly related to the contrast. 
Understanding where a filamentary structure merges into the surrounding background is the main critical point to be addressed, because it not only determines the extent of the region but also allows a realistic estimate of the background without which a reliable determination of the properties of the filament is difficult to obtain.

We present here a  method to detect and extract complex filamentary structures of variable intensity from 2D maps in the presence
of high and variable background. In section \ref{method} we describe our methods, in section \ref{tests} we apply the algorithm to realistic simulations of filaments superimposed on real observed background fields, proving the strength and the reliability of the method to identify filaments. Finally, in section \ref{SecDisc} we apply the method to real data, extracting physical parameters of the filaments, and we list our main conclusions in section \ref{SecCon}.


\section{Identifying filaments: The methodology}
\label{method}

Differential methods have already been proved to be useful to highlight structures like compact sources, e.g. the photometry code \textsc{CuTEx} \citep{molinari2011}, or filaments \citep[see Fig.\,3 of ][]{molinari2010}. In \textsc{CuTEx} the multidirectional second derivatives  are used to enhance the portion of the map with the strongest curvature of the intensity field along four fixed directions ({\it x}, {\it y} and the diagonals), corresponding to  the compact source centers due to their particular symmetries. \citet{molinari2011} have shown that the same operators qualitatively also trace the edges of extended structures like filaments. However, unlike sources, filaments are not strongly highlighted in the derivative along the four directions adopted by \textsc{CuTEx}. Therefore, we generalized the approach adopted with \textsc{CuTEx} by \citet{molinari2011} to the specific case of filaments.
To such aim, we initially follow the prescription described by \citet{Bond2010b} for the classification of features, like filaments, voids and walls, present in smoothed galaxy distributions. In their work, the authors noticed that each feature has a particular ``fingerprint" of the curvature along the principal axis. For example, on first approximation, filaments can be considered as cylinder-like patterns that are more convex along one direction with respect to the orthogonal one; in particular, the difference in curvature between the two directions would be the highest if the chosen directions are along the cylinder axis, which would have a flat curvature) and the orthogonal radially directed one, that would have the larger convexity. As the curvature of an intensity map along any direction is proportional to its directional second derivative, the Hessian operator is most suited to characterize the spatial properties. 
Differently than the fixed directions adopted for \textsc{CuTEx}, the eigendecomposition of the Hessian matrix, {\it H(x,y)}, of the emission intensity field {\it I(x,y)} (dust thermal emission in the far infrared in this particular case) immediately gives  the directions of the principal axes at each position {\it (x, y)} of the observed map, by means of the two eigenvectors, $A_1$ and $A_2$. The two eigenvalues \lone\ and \ltwo\ are proportional to the curvature at {\it (x,y)} along these direction. As we are focusing on the detection of emission features, we will be interested in convex morphologies, i.e. \lone $\le$ \ltwo $\le$0. In this notation we will assume that direction 1, being the one of maximum absolute curvature, will identify the cross-filament direction. A simple analysis of these eigenvalues can in principle give the direction, shape and the contrast of the local structure. For the case of a filament and {\it near its axis} the relationship:

\begin{equation}
\lambda_1 \ll \lambda_2 \le 0 
\label{bondcrit}
\end{equation}

\indent should hold, with the filament axis defined by the direction of $A_2$. Although useful for tracing the features on simulations, or relatively smooth data, this approach has some drawbacks when applied to maps of the interstellar medium. In fact, the above relation does not hold close to strong overdensities like, for example, compact clumps or cores found along the filament. In these cases \lone $\simeq$ \ltwo $\leq 0$ and, therefore, it is not possible to define the principal directions with enough accuracy. Moreover, equation \ref{bondcrit} holds only near the ridge of the filament, with the predominance of one eigenvalue with respect to the other weakening as one moves radially away from the filament center. Although we have to relax the formal criteria in Eq.~\ref{bondcrit} to identify filaments, maps of the second derivatives have the advantage that they filter out the large scale emission and emphasize the more concentrated emission from compact sources and filaments \citep{molinari2010,molinari2011}, due to its ability to pinpoint  strong variation in the gradient (i.e. change in the contrast) of the intensity distribution. Hence, whole filamentary regions, and not only their axis (hereafter ``spine") are included in the regions defined by a simple, conveniently chosen, thresholding of the second derivative map.

The pixel-to-pixel noise has a strong impact on the spatial regularity of the Hessian matrix, even for a relatively high signal to noise (S/N) map. In fact, the noise is amplified in the {\it H(x,y)} by the derivative filter, that is by construction a high-frequency passband filter, and then it affects the estimation of the correct local eigenvalues  \lone and \ltwo . The amplitude of the increase of the noise depends on how the differentiation is implemented. For the case of a 5-point derivative, see formula (3) of \citet{molinari2011}, we estimate an increase in the noise level of $\sim$20\% on the second derivative images and a further increase of $\sim$15\% in the eigenvalue maps. Smoothing reduces  the noise, but it also blurs the map, damping the variations on the small spatial scales and hence the contrast of the filament. Our tests indicate that smoothing through a gaussian with HWHM of the order of an instrumental beam represents a reasonable compromise between the need for noise suppression and blurring of the structure. With such a choise the pixel-to-pixel noise is reduced roughly by a factor $\sim$2 while the variation in the contrast decreases at the most by 20\%  with respect to the unsmoothed value. 

Thus, for a given intensity map we then compute the Hessian matrix, diagonalize and sort the eigenvalues in each pixel, producing two maps of eigenvalues \lone (x,y) $\leq$ \ltwo (x,y). We exclude from the analysis the pixels where \lone(x,y) $\geq$ 0, which identify concave shapes in the emission map. Two possibilities can occur for the remaining pixels: \ltwo (x,y) $<$ 0, that identifies convex regions, or \ltwo (x,y) $\geq 0$ for saddle points. Both cases occur in typical filamentary features with modulated emission along the axis. 

The next step is to threshold the eigenvalue map with the highest absolute value of \lone (x,y). The adopted threshold defines the lowest contrast that a region should exhibit to be considered as belonging to a filamentary structure. The optimal choice of the threshold depends on the condition of the map on which the user is working; in particular it depends on the strength of the diffuse background emission and of the pixel-to-pixel noise. 

We apply a morphological closing operator with a structural element half as wide as the beam to smooth the edges of the identified regions on scales smaller than the beam,  similar to what was done by  \cite{Rosolowsky2010}. Then, we proceed by identifying connected regions of thresholded image pixels and label them by progressive numbers; these regions are called Regions of Interest (RoI, hereafter). The border of each RoI represents a first rough estimate for the edges of the filament. However, since we use relaxed criteria with respect to \citet{Bond2010b}, different types of structures might contaminate the sample of candidate filaments. In particular relatively roundish structures like large and elongated compact clumps, or clusters of compact objects lying on a strong intensity field, might be selected as well. To remove this contamination we carry out an ellipse fit to each candidate RoI and discard all regions for which the axis ratio of the fitted ellipse is above a fiducial value of 0.75. Additionally, we also require that the major axis has a minimum length of three times the instrumental point-spread function of the image to exclude slightly elongated sources that cannot be considered as filaments. However, it is observed that there may be cases where filaments may intersect and generate a web-like structure that, depending on the contrast threshold adopted, may be catalogued as one single region; if the overall shape of the RoI happens to be more or less roundish, it would be discarded by the above criteria. To prevent this from happening, we also compute the filling factor of the RoI, as the ratio between the area of the RoI and the area of the fitted ellipse; regions whose filling factor is less than a fiducial value of 0.8 are kept as candidate filamentary structures. The fiducial ellipticity and filling factor threshold values adopted to identify and discard ``roundish" clump structures have been determined from tests carried out on Hi-GAL maps; the values can, however, be modified as an input to the detection code.

\begin{figure} 
\includegraphics[height=\columnwidth,angle=0]{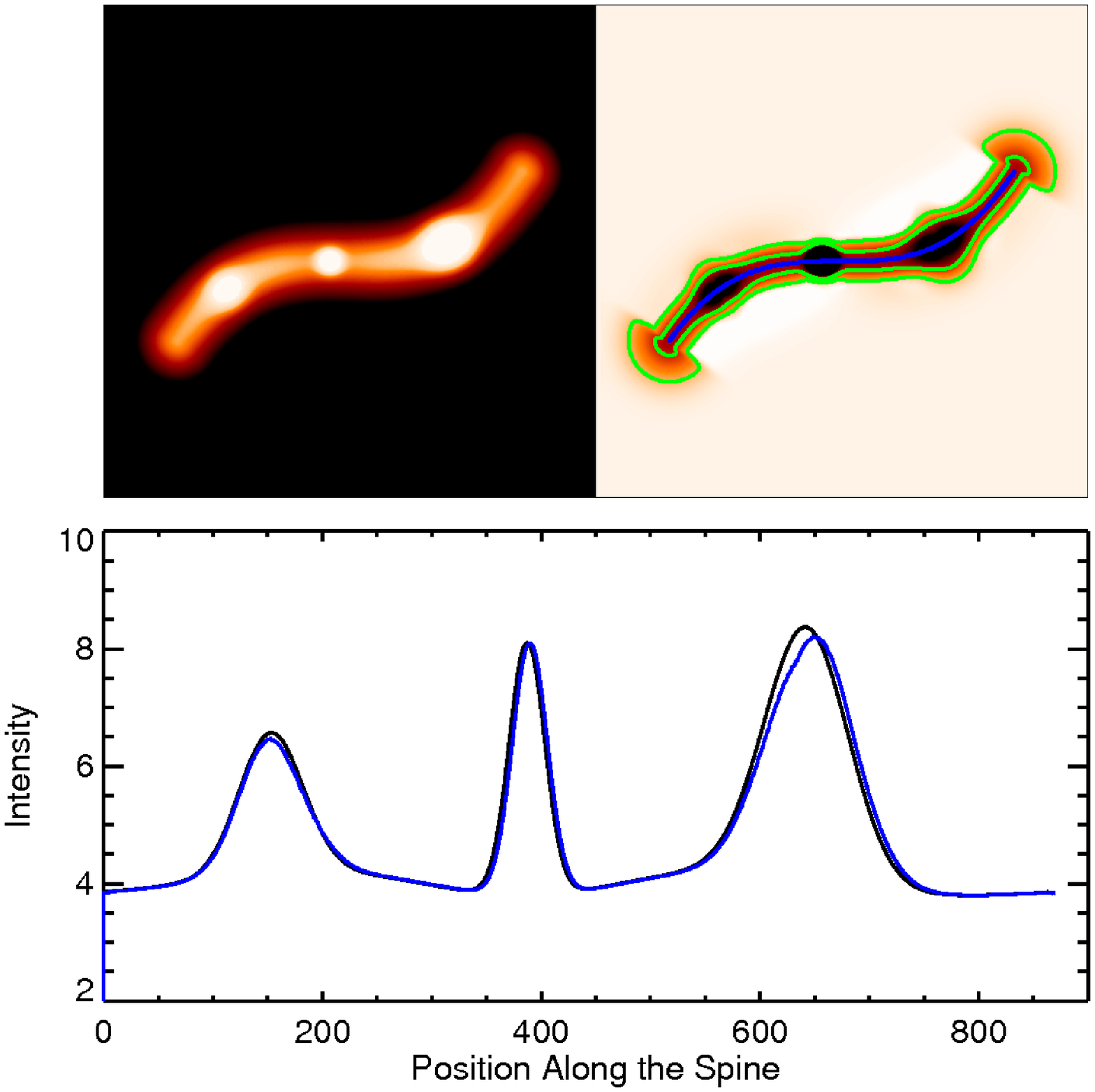}
\caption{Example of our method applied to a single filament (top-left panel) with a 20\% flux modulation and three bright clumps distributed along its axis. The top-right panel shows the image of the minimum eigenvalues of the Hessian matrix; the green contours mark the region of interest resulting form the thresholding at two different levels, a high and a low threshold,  while the blue line represent the estimate of the spine of the filament after the "thinning" of the RoI. In the lower panel are shown the simulated filament spine profile (black) and the extracted profile for the low threshold (blue).}
\label{onefil}
\end{figure}

Once the list of candidate filamentary RoIs has been decontaminated from compact roundish clumps or undersized elongated structures, we proceed to identify the spine of the filament by applying a morphological operator of ``thinning''  on each region (see \citealt{GW92}). In short,  a ``thinning'' operator on a RoI works by correlating each pixel and its surrounding with specific binary masks defining specifical patterns. These patterns are designed to determine if a pixel belongs to the RoIs boundary; in such a case the pixel is removed. We adopted a 3$\times$3 binary mask, so the classification of a pixel as a boundary depends strictly on its closest neighbours. The same approach has already been applied in a different field like the identification of filaments on the Sun \citep{Qu2005}.
Under the assumption that the filament is symmetric in its profile, by repeating the procedure iteratively until no further pixels can be removed, the surviving points constitute the spine of the filament. In the case of slightly elliptical blobs, the structure would reduce to few pixels, or even to one point in the circular case, that are filtered out when applying again our criteria on minimum length. The spine pixels are then connected  through a ``Minimum Spanning Tree'' (MST), implemented with the Prim algorithm \citep{primalgorith}, to define the unique path that joins them together. This allows us to identify nodal points where multiple branches depart and immediately enables the classification of structures in main hubs of peripheral branches. An example of the method applied over a very simple and bright simulated filament is given in Fig.~\ref{onefil}. The simulated filament has variable intensity along its spine with periodical fluctuation of amplitude equal to 20\% and few compact sources of different size distributed along its axis; although this seems an idealized situation, the presence of the three sources would have caused the original method of \citet{Bond2010b} to break the filament into three portions.

\begin{figure} 
\includegraphics[width=8cm]{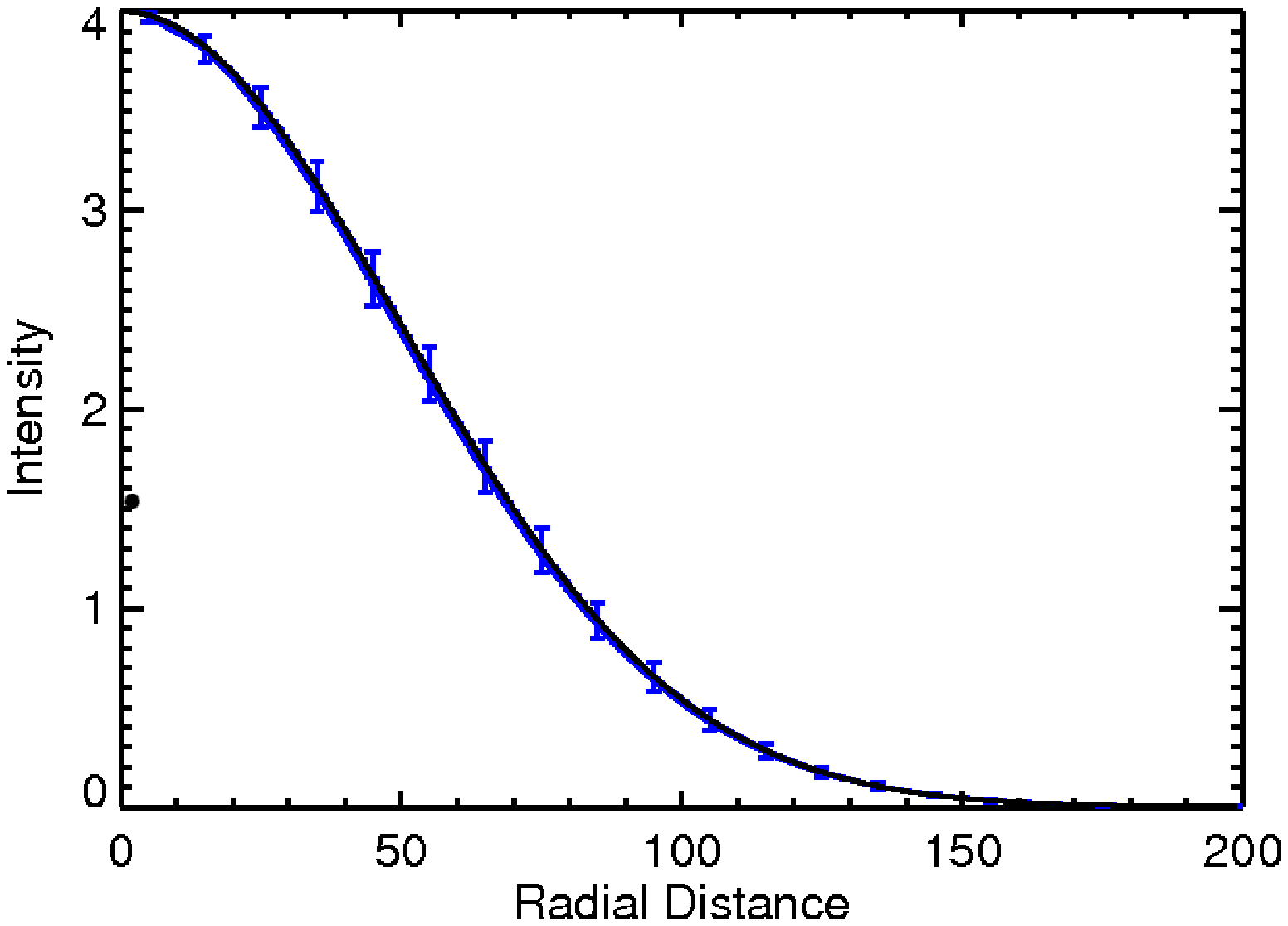}
\includegraphics[width=8cm]{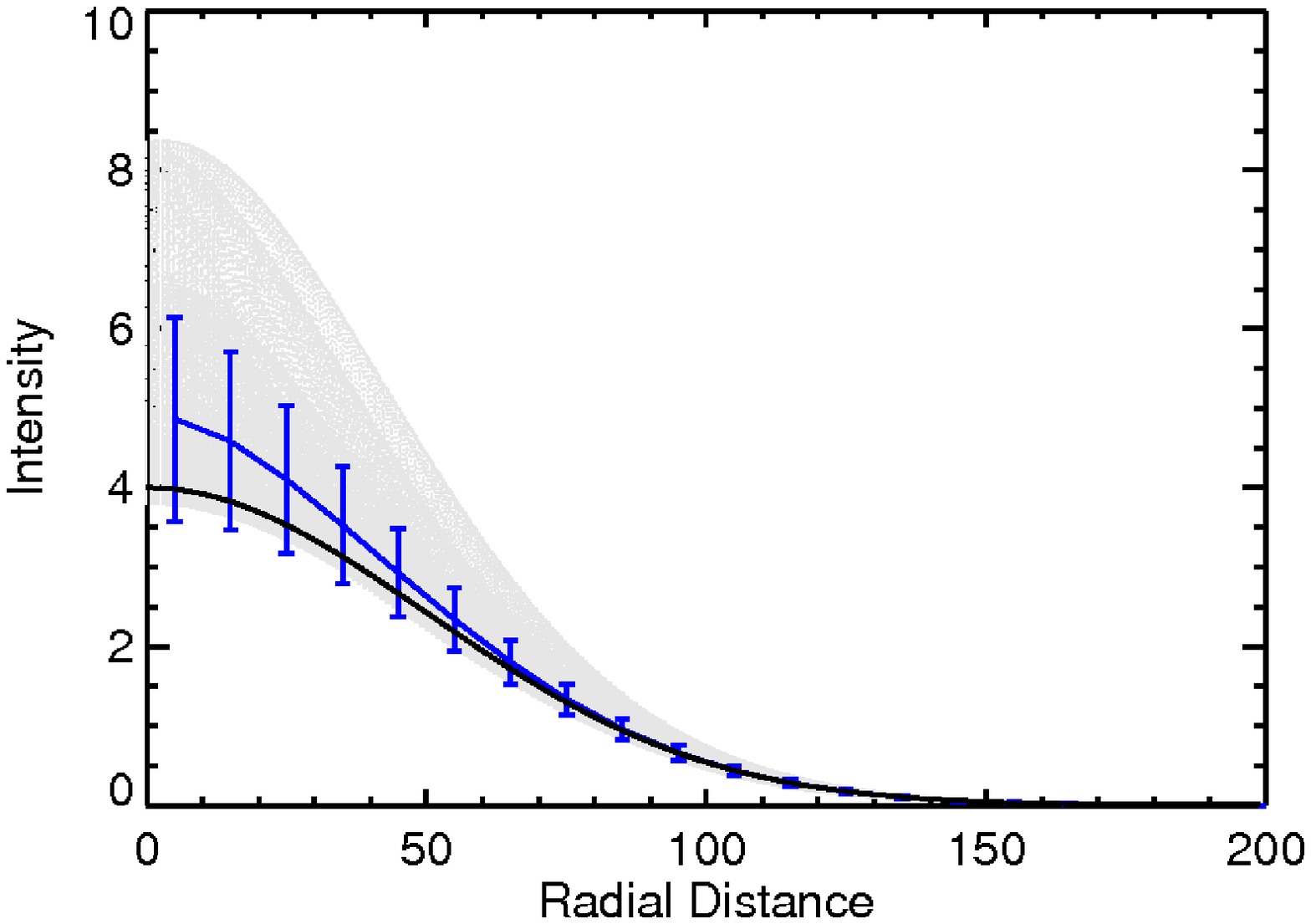}
\caption{Example of the ability to retrieve the width of a simulated filament. In the left panel we see that the method retrieves (blue line) exactly the cross-spine width of the filament in input (black line) when the filament has no sources along its axis. In the right panel we show the results for the same filament in Fig. \ref{onefil}; the ensemble of the cross spine profiles (the set of grey points) shows more dispersion with respect to the width of the input filament (black line). The fit to these points (the blue line) represents the average filament cross-spine profile and it is overestimated with respect to the true value (although consistent within the uncertainty bars that reflect the spread of filament width along the spine).}
\label{onewidth}
\end{figure}

Our main goal in this exercise is to obtain the physical characterization of the filamentary structures. For example, we would be interested not only in knowing where filament spines are, but also what their masses are. To this aim, we also need to estimate the cross-spine size of the filament as well as of the underlying background level that we need to subtract to obtain the true contribution to the emission of the filament material alone. To do so, for each spine point we fit the brightness profile in the direction orthogonal to the spine with a Gaussian function and compute the median of all the FWHM values obtained; an associated uncertainty is provided by the standard deviation of the individual width estimates along the filament (see Fig.~\ref{onewidth}). A new region mask is then created, symmetrical around the spine and with total width equal to twice the median filament FWHM. For very complex features where filaments are organized in web-like structures, the cross-spine profile fitting often fails to converge as there are not enough background pixels to reliably constrain the fit. 

We then provide an additional measure of the filament width by adopting the initial RoIs identified by the Hessian eigenvalues thresholding, and enlarging them with the morphological ``dilation" operator  \citep{GW92} applied three times in sequence. The merit in doing this is that whatever the threshold adopted, the thresholding is always done over the map of minimum \textit{but negative} eigenvalues; in other words, the pixels selected will always belong to regions where the curvature of the brightness profile in the maximum curvature direction is within the convexity region. This implies in general a conservative identification of the filament region, because it would neglect the wings of the filaments where the emission profile changes concavity before joining the background emission. In cases of isolated filaments where the cross-spine Gaussian fitting converges, we will show (see below) that the two width estimates are in very good agreement.

The measurement of the background level on which the filamentary structure is sitting is very important for a reliable measurement of the total emission of the filament, or of the total mass in case the filament detection is run on a column density map as we will show below. Once the filament RoI and spine have been determined, we proceed as follows. For each spine point we compute the direction perpendicular to the local spine, and we select the pixels intersecting the 2-pixels-wide boundary region surrounding the filament RoI on each side of the spine along such a direction. These two sets of pixels (one on each side of the filament) are fitted with a line producing a reliable local estimate for the background. This is repeated for all filament spine points, producing a detailed estimate of how the background varies along the filament. 


\section{Code performances on simulated and real filaments}
\label{tests}

To characterize the performance of the software in conditions that more closely resemble real situations, we carried out an extensive series of tests using sets of simulated filamentary structures superimposed on maps of the ISM emission showing a strongly variable background. As it is best to test in the most realistic conditions possible, we used maps from the Hi-GAL survey.

While the simplest and ideal shape of a filament is an homogeneous straight cylinder shape, those conditions are rarely (if ever) found in observations. Gravity, turbulence and other  evolutionary effects twist the orientation of the ``ideal'' shape and also gather the matter in different places along the original structure. Moreover, what we generally see is a 2D projection of a 3D structure that, depending on the viewing angle, may significantly amplify any departure from the ideal elongated cylinder shape. A realistic filament is generated by first computing randomly twisted curves as the ``spine'' for the structure, with variable profile along such curves up to 20\% with respect to the mean intensity . The brightness distribution in the cross-spine direction is assumed to be Gaussian, and in some cases we added compact sources of different sizes at different positions along the spine. Filaments were then randomly rotated to avoid any bias from specific orientations. Regridding of the simulation has been also included to estimate the effect of the pixelization on the performance of the method. We produced various set of simulations, with different numbers of filaments and degree of clustering, divided in two groups defined by the width of the structure simulated: unresolved, ''thin'',  and resolved, ''thick'', filaments.

We want to stress here that our method is suited to identify extended, but still concentrated, emission. The emission from the large scale structures is strongly dampened in the derivative maps, so broad filaments can be detected only if they have strong central intensities. In fact, the dampening in the derivative map increases with the spatial scales following a power-law behaviour with an exponent of 2 for scales $\ge$6 pixels, i.e. 2 times the PSF for Nyquist sampled maps, see also  \cite{Molinari2014}. Therefore, for example, structures with a typical scale of $\sim$2.5 times the PSF have their central intensities dampened by a factor of $\sim$10, instead the ones with scales of $\sim$7.5 times the PSF will appear $\sim$100 times fainter in the derivative map than in the original intensity map. Hence, for a fixed threhold level on the same background, it is possible to identify, if they exist, structures with scales of the order of $\sim$2.5 times the PSF and intensities 10 times fainter than the ones with scales of $\sim$7.5 times the beam. It is clear that the closer the width of the structure will be to the scales of the background emission, the harder the structure will be distinguishable. They will stand out on the derivative images only if their intensities are comparable with that of the smallest scales present in the background component. 

\subsection{Unresolved simulated filaments}

\begin{figure} 
\centering
\includegraphics[width=\columnwidth,angle=0]{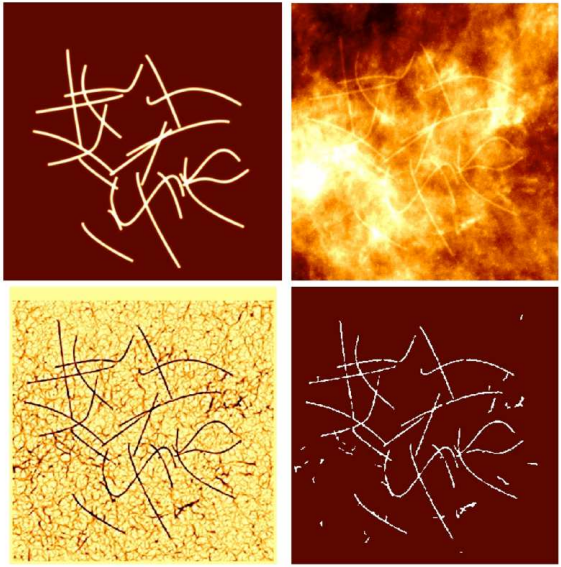}
\caption{Results of the filament detection over a set of 25 simulated filaments with FWHM of 3 pixels in the top-left panel, overlayed on a patch of emission from Hi-GAL 250\,\um\ (top-right panel) used as a background. Only the highest contrast filament/background  situation simulated is reported in the figures. The bottom-left panel shows the minimum eigenvalues image used for thresholding, while the bottom-right shows the masks of the recovered filamentary structures; this shows more output filaments than the 25 input filaments
because the background image used is real {\em Herschel} data from the Hi-GAL survey and as such contains real filaments.}
\label{simulthin}
\end{figure}

In Fig.~\ref{simulthin} we show an example of a simulation: the presented case has 25 filaments all having a cross-spine size with FWHM of $\sim$3 pixels, namely the size of 1 PSF, corresponding essentially to unresolved filaments assuming a fully Nyquist sampled map, like the Hi-GAL maps \citep{molinari2010}. The set of filaments was distributed over three different patches of diffuse emission extracted from Hi-GAL 250\,\um\ maps to try and make results independent from a specific local background condition. For each case, we normalized the mean intensity along the spine of the simulated filaments to the median value of the background in the patch to achieve a contrast level equal to 1. Moreover, we also generated images where the brightness of the filaments was decreased by a factor of 2 and 4 with respect to the background image, to simulate filaments with different contrast levels (as an example we present the simulation with contrast 0.5 in Fig.~\ref{simulthin}, top right). The filament extraction method is then applied over all simulated fields (25 filaments, for three different background configurations, for 3 different filament/background contrast ratio), using 4 different extraction thresholds. 
The code performances are characterized by comparing the length, width and area of the recovered filaments with those of the input simulated ones. 

\begin{figure} 
\includegraphics[width=\columnwidth,angle=0]{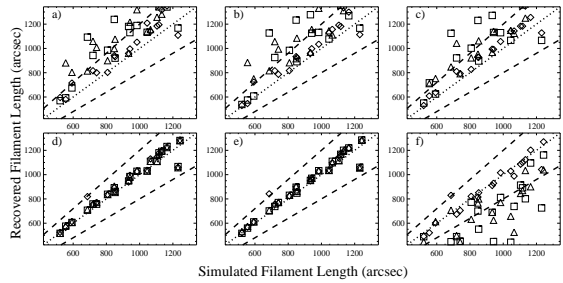}
\caption{Results for the recovered filaments lengths in the case of unresolved filaments versusf the true length. Each symbol represents one filament, and the three different symbol types are for the three background types (see Fig.~\ref{simulthin}). The top row is for the lowest used extraction threshold, which is more sensitive to relatively fainter structures; the bottom row is for the highest used threshold, which is less sensitive to fainter emission. For each row, the three panels  show the results for the three filament/background contrast ratios used: in the left panel is the nominal situation that is represented in Fig.~\ref{simulthin}, the center panel is for contrast reduced by a factor 2, while the right panel is for contrast reduced by a factor 4. In all panels the dotted line represents the identity line, while the dashed lines indicate a 20\% discrepancy.}
\label{res-thin-length}
\end{figure}

Figure~\ref{res-thin-length} reports the results for the recovered filament length as a function of the input length. Results are shown for the lowest (top row) and the highest (bottom row)  extraction thresholds, and for decreasing filament/background contrasts (from left to right). On average the results are very good, with recovered length that in most situations agrees with the input values within 20\% for the range of thresholds adopted. In the case of the lower extraction threshold, we see a general trend to obtain lengths that are systematically overestimated by about 20\%\ irrespective of the filament contrast. This can be explained by the fact that with low thresholds on the minimum Hessian eigenvalue the regions initially selected by the thresholding are larger, and the subsequent ``thinning" systematically produces longer spines. The situation clearly improves going to higher thresholds for nominal and halved contrast ratios (Fig.~\ref{res-thin-length}\textit{d} and \textit{e}) independent of the type of background used; if, however, the higher thresholds are used and the contrast gets too low (panel \textit{f}, a factor 4 less with respect to the situation depicted in Fig.~\ref{simulthin}), then the code starts to break the filaments up into shorter portions depending on the background where the simulated filament falls. However, for intermediate thresholds   lengths of the structures are still recovered within 20\% accuracy even for this faint case.

The code behavior in recovering the average width of the filaments is more regular, independent of the extraction threshold, background type and contrast: the recovered widths mostly agree better than 20\%\ with respect to the input values. 

\begin{figure} 
\centering
\includegraphics[width=\columnwidth]{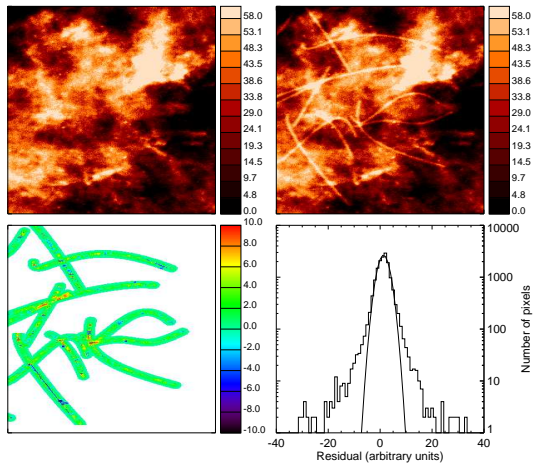}
\caption{A typical field of the Galactic Plane at 250~\um\ (top-left), with a superimposed set of simulated filaments(top-right); the color-scale provides the absolute signal levels of the map in arbitrary units. The simulated filaments shown here have a mean intensity value along the spine equal to 18 arbitrary units. The bottom-left panel shows the difference between the background estimates at the filament positions  and the original background. The bottom-right panel shows the intensity distribution of the differences of the bottom-left panel only for the pixels belonging to filaments. In red line we overlap the gaussian fit to such distribution.}
\label{backver}
\end{figure}

An accurate estimate of the filament background is critical for a reliable measurement of the intrinsic filament emission, whether it is total flux or total column density (if run on a column density map). In Fig.\,\ref{backver} we show an example of the reliability of our background estimates. We take a real field over the Galactic Plane (top-left) and superimpose a set of simulated filaments. We normalized the filaments to have constrast 1, 0.5 and 0.25, (only the case with contrast 0.5 is  shown in the top-right panel). For the patch of background presented in Fig.\,\ref{backver} the mean  intensity along the filament spine is 35, 18 and 9 arbitrary units respectively for contrast 1, 0.5 and 0.25. 
We then extract the filamentary structures, identify the ones that correspond to the simulated filaments (these are the only ones for which we have a truth table) and subtract only them. In the bottom-left panel we show the difference between our estimate of the background after the subtraction of the filaments and the initial background (top-left panel). The distribution of such differences between the input and the filament-subtracted backgrounds, for all the pixels where a simulated filament was inserted, is shown in the bottom-right of Fig.~\ref{backver} with a gaussian fit overplotted in red. The distribution is centered around zero difference, with 95\% of the pixels falling in the gaussian fit with a FWHM of 5 in arbitrary units. The remaining 5\% of pixels shows residuals as large as 30 arbitrary units and are generally located at the bright position of the original background map (with values as large as 80 arbitrary units), sometimes on real compact sources, or where multiple filaments nest each other. Very similar distributions are found for all the contrast cases and depends mostly on the background, with residuals generally small with respect to the distribution of background intensities at the filament position.
In other words, the code delivers reliable estimates of the filament underlying backgrounds.

\subsection{Resolved simulated filaments}

\begin{figure} 
\includegraphics[height=\columnwidth,angle=-90]{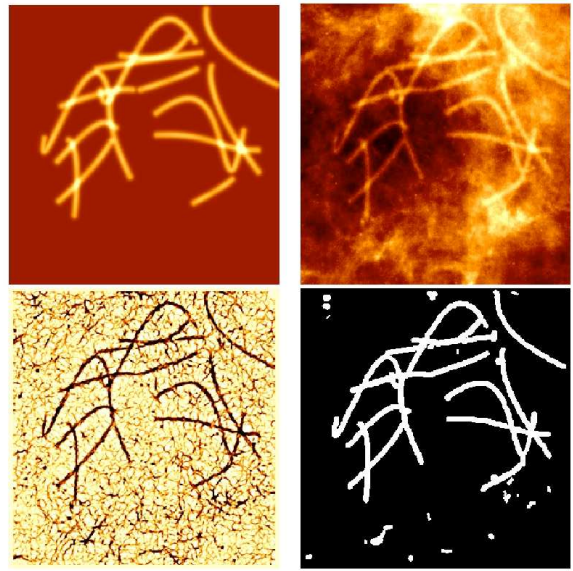}
\caption{Same as Fig. \ref{simulthin}, but with filaments three times wider and differently distributed on the background image.}
\label{simulthick}
\end{figure}

We further carried out simulations where the filaments are assumed to be resolved. Fig.~\ref{simulthick} is the analogue of Fig.~\ref{simulthin} for a different filament distribution, but in this case the filaments have a FWHM three times larger. As the intrinsic curvature will be lower for these extended structures, we expect the code performance to degrade accordingly. In fact, while the background has not been changed, the filaments have a shallower intensity variation along the radial direction, so they are less prominent in the derivative image.

\begin{figure} 
\centering
\includegraphics[width=\columnwidth]{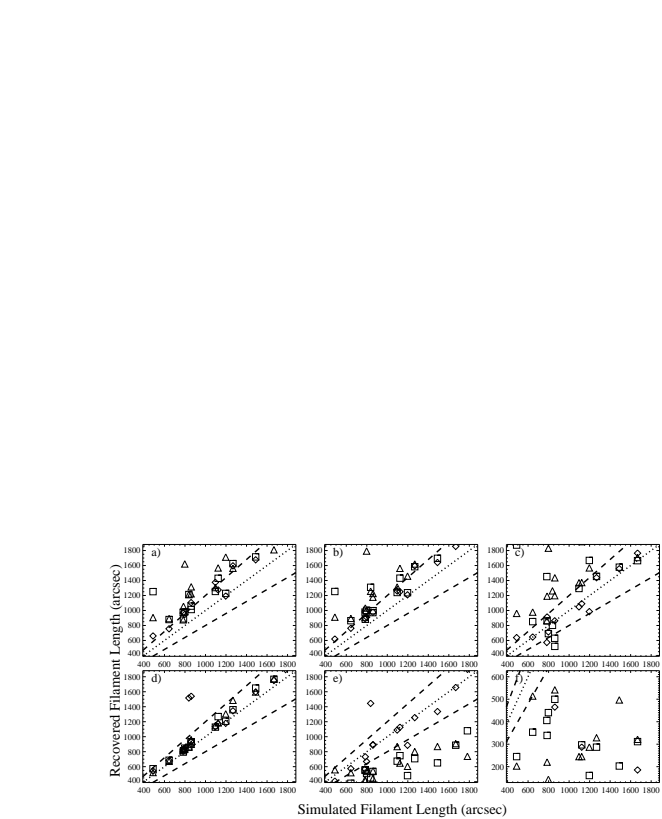}
\caption{Same of Fig. \ref{res-thin-length}, but for the case of resolved filaments.}
\label{res-thick-length}
\end{figure}

The recovered lengths for the retrieved filaments are shown in Fig.~\ref{res-thick-length}, where the meaning of the symbols and of  the different panels is the same as in Fig.~\ref{res-thin-length}. The code continues to perform very well on average, showing very similar behavior as for the unresolved features, up to moderate contrast between filament and background. 
For the lowest contrasted filaments simulated here, the code has more trouble recovering the correct length for low detection thresholds, producing a larger scatter of values (Fig.~\ref{res-thick-length}\textit{c}) than the unresolved case. In addition, at higher thresholds, the moderate and low contrast filaments are also undetected, and the few detected ones are broken into smaller portions (Fig.~\ref{res-thick-length}\textit{f}). As expected, therefore, this method based on the second derivatives that are computed over a discrete set of pixels performs less and less reliably the shallower the structure. It is fair to point out that this break-down in performance is experienced for very unfavorable conditions where the filament/background contrast is 4 times less than what appears in Fig.~\ref{simulthick}. If the contrast is decreased by only a factor of 2 (Fig.~\ref{res-thick-length}\textit{b} and \textit{e}) the code performs much better in recovering the length. 

The situation is worse when one considers the widths of the filaments for the resolved case. As the features are much shallower than the unresolved filament case, while the spatial dynamical range of the background has not changed, the Gaussian fits performed at all spine positions are less constrained. The average result is that in the best contrast situations the width is systematically  underestimated by about 20\%. For lower contrast filaments the determination is much more noisy and the width is recovered with an uncertainty of the order of 30-40\%. 

\subsection{Filaments widths and background estimates on a real filament}
\label{realestim}

\begin{figure} 
\includegraphics[width=9cm]{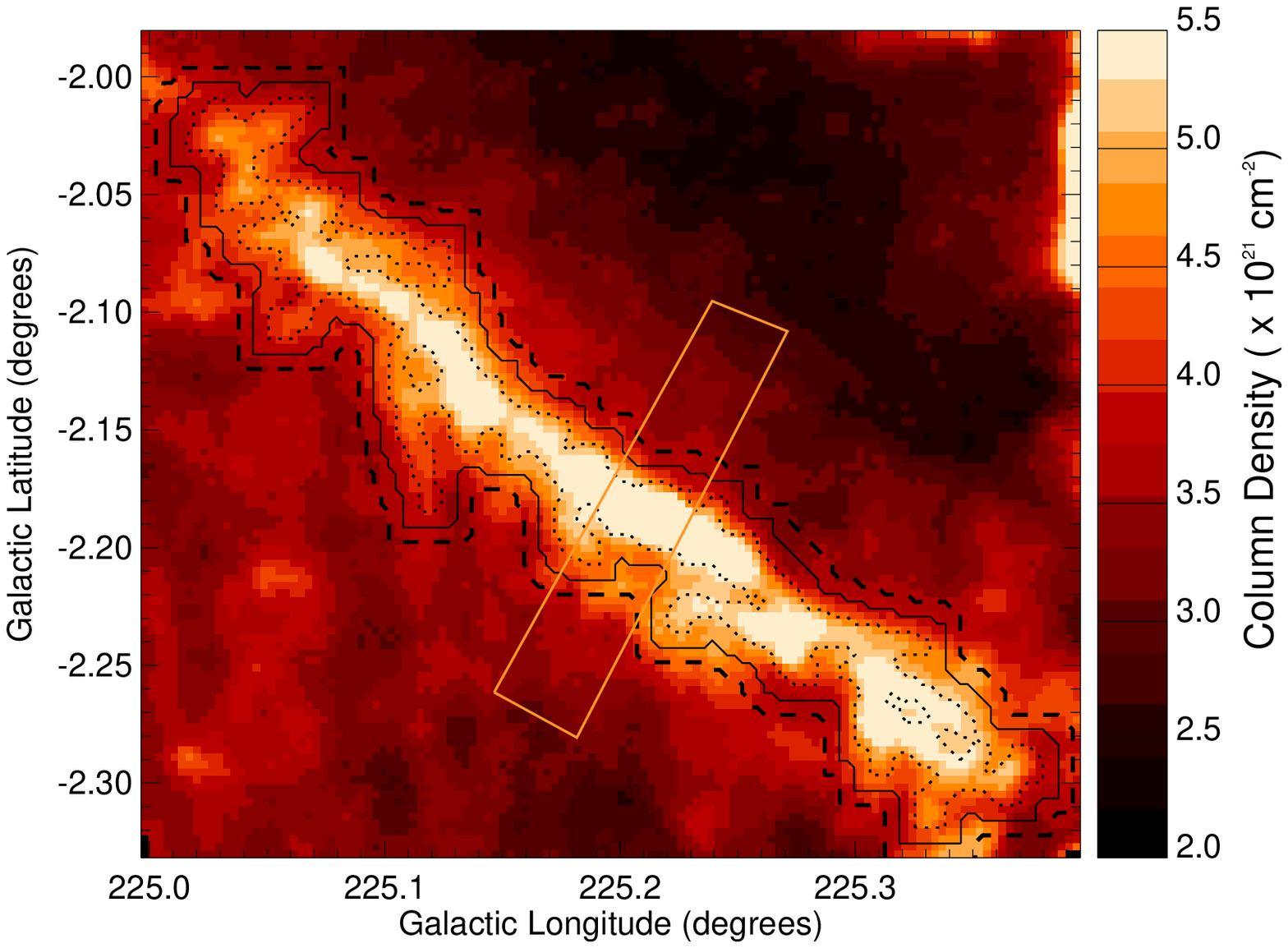}
\includegraphics[width=9cm]{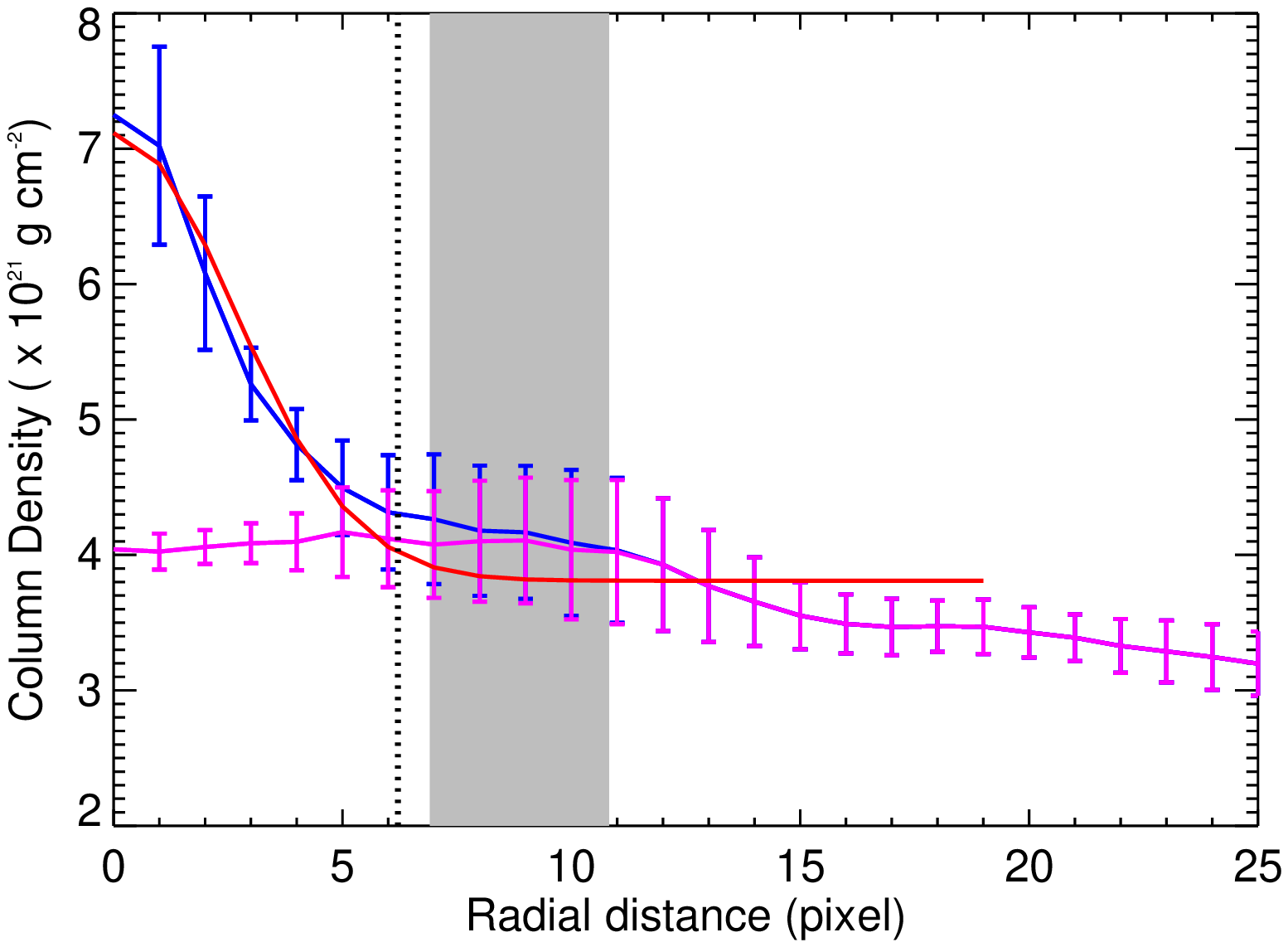}
\caption{
\textit{Top:} 
subsection of the larger column density map of \cite{elia13} with a typical filament. The dotted line encloses the RoI selected by the initial thresholding on the Hessian eigenvalues; the full black line is the RoI after applying the dilation operator three times; the dashed line encloses the area over which the background is estimated. 
\textit{Bottom:} cross-spine radial profile of the column density (blue line) for the region enclosed in the yellow-line rectangular area in the top panel. The full red line is the Gaussian fit, with the dotted vertical line marking a typical radial distance of 1 FWHM from the central spine. The full magenta line is the background profile estimated from the column density corresponding to radial distances falling into the shaded area.}
\label{real_filwidth}
\end{figure}

To prove in more detail the ability of our approach to recover a correct estimate of the filament width and background level, we illustrate the algorithm performance results on a real filament extracted from the more general results that will be presented in the following section. Fig.\,\ref{real_filwidth} shows a typical real situation for a relatively isolated filament. We see that the average width of the filament as it would be estimated from the Gaussian fitting of the radial profile as explained in \S\ref{method} (the dotted vertical line in Fig.\,\ref{real_filwidth}-bottom panel) is in excellent agreement with the cross-spine size of the final filament RoI (after applying the dilation operator), that corresponds to the left boundary of the grey shaded area in Fig.\,\ref{real_filwidth} (bottom panel), and to the black line in Fig.\,\ref{real_filwidth} (top panel). The shaded area corresponds to the radial distance spanned by the pixels that in Fig.\,\ref{real_filwidth} (top panel) are enclosed between the full and dashed black lines, where the background is estimated. 

We point out that our method is totally consistent with classifing  {\it as filamentary} all the pixels within the borders defined by the flattening of the radial profile.  Such definition has already been used in previous work on filaments, i.e. \citet{Hennemann2012}.

\subsection{Performance evaluation}
\label{performance}

The results of the extended set of simulations illustrated in the previous sections build confidence in the filament extraction method that we have developed. The method has been proved to identify easily structures that are as large as 3 times hypothetical spatial resolution element of the test maps. 
It is clear that the best situation is for unresolved filaments, where the curvature of the brightness distribution is higher. Filaments lengths and widths are in general recovered to within a 20\% uncertainty with respect to input values, unless the filament/background contrast is very low. As expected the situation gets worse when shallower filaments are used in the simulations. While these fainter structures are still identified, the estimation of their morphological and physical parameters become unreliable. Similar uncertainities arise for wider simulated filaments but with intensities comparable with that of the background component.  Structures wider than 3 times the PSF are identified only if they are relatively bright with respect to the background. In such a case the estimation of the parameters is satisfactory due to  the high contrast filament/background.

To summarize, the output of method is very reliable for structures as wide as 3 times the spatial resolution element of the map. The performances quickly degrade for wider structures which, due to the intrinsic degeneracy in the method between the width  and the central intensities,  can be identified only if they are as brigth as the background. For a fixed threshold, the widest structure that can be identified depends on the background properties (its smallest scale and the relative intensity).

Another relevant point coming out of the simulations is that different thresholds are appropriate to highlight different kinds of structures, resolved/unresolved and with different contrast intensity, over different background values. While a high threshold value is able to properly recover unresolved filaments with high/moderate contrast with respect to the background, it splits faint structures into multiple segments of shorter lengths. However, adopting a low threshold enlarges the identified RoIs and artificially increases the filament lengths in the high/moderate case. 
Ideally, we would like to apply a higher threshold on regions where the filament variations dominate over the background  and a lower threshold where they are shallower and fainter. 

Hence, we adopt as a local estimator for the threshold the standard deviation of the minimum eigenvalue computed on map regions 61$\times$61 pixels wide.  Increasing the region size does not change substantially the minimum threhsold value. This is expected since by enlarging the region where the threshold is computed, we are including the contribution from larger scales, which is neglegible for scales greater than $\sim$60 pixels. In fact, those scales are dampened up to $\le$0.5\% of their original value. 
With such a choice, the threshold will be higher in regions with large and intense fluctuations of the emission, eventually dominated by the presence of the filaments, while it will decrease in detecting regions with a shallower contrast where the variations are smaller.
 
Finally, it is worth noticing that while it is straightforward to identify filaments as elongated structures in the isolated cases, it clearly becomes difficult when multiple objects overlap each other like in the cases of our simulation. On real data, multiple filaments can be physically connected and converge toward larger structures, called ``hubs'' \citep{Myers2009} or crossing each other due to line of sight effects. From this point on we will  call {\it one filament} a whole region corresponding to one identified RoI. However, in the case of complex RoIs the axis is not a simple segment, but often it can be composed of multiple segments connected to each other in nodal positions. We will call each one of those segments a {\it branch}. 
These branches reflect asymmetries of the RoI and they have two different physical interpretations: {\it a}) they represent the portion of a larger filament between two local overdensities inside the structure, {\it b}) they are physically separated filaments, connected to the main structure by our algorithm since there is not a strong discontinuity in the contrast variation.

\begin{figure} 
\includegraphics[width=\columnwidth]{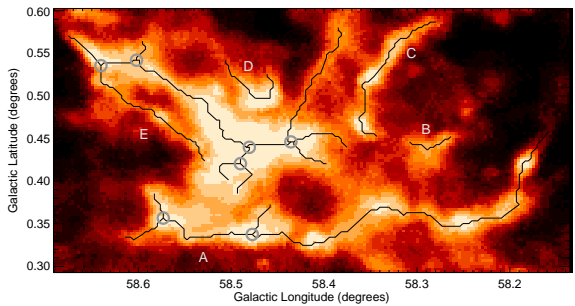}
\caption{Column density map of a portion of the Hi-GAL observation centered at  {\it (l,b)} = (58.3917, 0.4235) and wide 32' $\times$ 17'. The axis of five filamentary regions, indicated by the letters from {\it A} to {\it E}, are shown as thick black lines. Filaments {\it A} and {\it E} are composed respectively by 5 and 12 branches connected to nodal points (indicated by grey circles in the figure).}
\label{ExpBranches}
\end{figure}

As an example in Fig.~\ref{ExpBranches} we show a 32'$\times$17' wide region of the column density map centered at {\it (l,b)} = (58.3917, 0.4235) computed from {\it Herschel} Hi-GAL observations. In this figure there are five filaments, only three of which, indicated by {\it B, C, D}, are composed of a single axis. The two remaining, identified as {\it A} and {\it E}, are composed by multiple branches starting from the nodal points, indicated with a grey circles in the figure, and trace the main remaining structures. 

The physical quantities for each branch, like mass or column densities, are  computed on the subregions of the original filament mask determined by associating with each branch all the pixels that are closer to the relative branch spine. 

\newpage
\section{Discussion of the filamentary structures in the Outer Galaxy}
\label{SecDisc}

Our filament identification algorithm was run on our column density map which was calculated from the four {\em Herschel} Hi-GAL maps in the Galactic longitude range of $l$=216.5$^{\rm o}$ to $l$=225.5$^{\rm o}$, hereafter indicated as $l$217--224, at the wavelengths 160, 250, 350 and 500\,$\mu$m. These maps are the first Outer Galaxy tiles published from the Hi-GAL survey. The $l$217--224 $70$\,\um\ to $500$\,\um\ and column density maps have been presented in \citet{elia13}, along with a compact source catalog. The adopted dust opacity law was $k_{0}({{\nu} \over {\nu_{0}}})^{\beta}$ with $k_{0}$ = 0.1\,cm$^{2}$\,g$^{-1}$ at $\nu_{0}$ = 1250 GHz (250\,$\mu$m) \citep{Hildebrand1983} and\,  $\beta$ was assumed to be 2. It is important to notice that the dust opacity parameters adopted to compute the column density map are rather uncertain. The column density values depend on the assumptions in the  dust emission model.
For example, the fixed spectral index $\beta$ = 2 might be wrong for the cold and denser regions, where a higher value for $\beta$ is expected. A lower value for $\beta$ influence the grey-body fit outputs with an overestimation of the temperature and an underestimation of the column density. Furthermore, a more realistic dust model can be adopted for the more diffuse material, see for example \cite{Compiegne2010}.  We estimate that the uncertainity on the dust emission model can affect our estimate of the column density map by a factor of $\sim$2.

\citet{elia13} determined the kinematic distances of the compact sources in $l$217--224 from the CO\,(1--0) emission observed with the NANTEN telescope. 
Clump distances range from 370\,pc up to 8.5\,kpc and, as shown in Figure 3 of \citet{elia13}, the degree of contamination from kinematically separated regions along the line of sight in the $l$217--224 longitude range is very low. 

The filament extraction was run with a threshold of three times the local standard deviation of the minimum eigenvalue (see Sec.\,\ref{performance}). Moreover, we filtered out any regions with a length smaller than 4 times the beam, i.e. $\sim$12 pixels or $\sim$2', as further constraint on the ellipticity of the structure, in addition to the one described in Sec.\,\ref{method}. With such a choice we are  excluding short, but more distant, structures. Hence, we expect that our sample will be incomplete in terms of wide angular sizes.  Moreover, even narrow structures can be missed by the detection algorithm, if their contrast variation is lower than the adopted threshold. Farther filaments have a shallower gradient of the contrast along their profile due to beam dilution. Thus, the sample will lack also of faint and narrow  sources whose variation is closer to the one of the background. Despite the incompleteness of the sample, our aims here are to give a first estimate of the statistical properties of structures that {\it look filamentary on Herschel maps}. The shorter and fainter structures are statistically represented in our sample by the nearby structures. However, we remark that since the region studied in this work is mostly dominated by the emission coming from distances less than 1.5\,kpc (Fig. 3 and Table 1 of \citet{elia13}) the incompleteness of the sample will have a minor impact on our results.

\subsection{Morphological properties of filaments}
\label{secfil}
The algorithm identified $\sim$500 filaments containing in total $\sim$2000 branches spread across the Galactic longitude range. The detected filaments are shown on the column density map in Figure\,\ref{OutputDetection}.  A visual inspection of the result indicates that all the major filaments identifiable by eye  are traced by the algorithms.

We cross correlated each filament RoI with the clump positions from \citet{elia13} and, for the filaments with a match, we assigned the distance given by the mean value of the clump distances found within their border. We found that 40\% of the detected filaments have at least one associated clump and their distances range between 500\,pc to 8.5\,kpc with a median distance of 1.1\,kpc (shown in Fig.~\ref{hist_filaments}, panel $a$) corresponding to the average distance of the CMa OB1 association \citep{ruprecht1966}. The distance distribution is compatible with the two main Galactic arm structures along this line of sight: the Orion spur locate at distance of $\le$1\,kpc and the Perseus arm at $\sim$ 2\,kpc. Few filaments might be associated with the Outer arm, located at a distance $\sim$5\,kpc, however we do not find a defined separation between filaments in such a structure and the one in the Perseus arm. The remaining 60\% of the filament sample lacking of a clear distance association were assumed to be at the 1.1\,kpc, i.e. the median of the distribution (not shown in panel $a$). In the remaining panels of Fig.~\ref{hist_filaments} we plot separately the filaments with a kinematic distance (i.e., filaments with clumps) and these without (i.e., filaments without clumps) in solid and dotted lines, respectively. We point out that the percentage of filaments without a clump detection inside their border is affected by the criteria adopted by \citet{elia13}. 
In fact, the catalog presented by these authors includes  the clumps identified on the Hi-GAL maps for which they could determine a distance estimation, through a detection in the NANTEN CO observations.
Hence, two effects contribute to the number of non detected clumps inside the filament:  the NANTEN observations {\it a}) do not cover the whole area surveyed by {\it Herschel} Hi-GAL data \citep[see Fig.\,4 of ][]{elia13}, and {\it b}) have a low sensitivity. We found that $\sim$10\% of the detected filaments fall outside the NANTEN coverage area. Furthermore, we compared the maximum column density found in each filament RoI and estimated that another 8\% of the sample are structures that might be undetected in the CO data. 

The histogram of the filament spine length (for those filaments with a kinematic distance) peaks at around 2\,pc, and despite the presence of a significant tail that extends up to 60\,pc, most of the filaments have lengths between 1.5\,pc and 9\,pc, with a median value of 2.45\,pc (panel $b$, solid black line). The filaments whose distance was assigned to the median value, show a different distribution (panel $b$, dashed black line): their lengths strongly peak at 1\,pc and then drop off quickly. The cut we have adopted in our selection criteria  translates to an artificial length-cutoff at 0.74\,pc for the median filament distance of 1.1\,kpc. The lengths are estimated from the map and are liable to projection effects due to a possible inclination effect. No information is available for possible inclination of these structures along the line of sight. Assuming a random uniform distribution for the inclination of the filaments, the observed mean value of the inclination angle with respect to the line of sight would be $\sim57^{\rm o}$, implying that the intrinsic filament length would be $\sim$19\% longer. However, due to projection effects we do not identify all the filaments that have a small angle between their axis and the line of sight, hence the true mean inclination would be larger. The net result is that the length distribution is closer to the intrinsic one.

In panel $c$ of Fig.\,\ref{hist_filaments} we show the distribution of the measured width for the identifed structures. Almost all the filaments are resolved in their radial direction (see Fig.\,\ref{Filwidth}) and only 8\% of the filaments with a reliable distance have widths that are compatible within the errors with the beam size. We stress that the majority of detected filaments have width $\sim$1.9 times the beam, despite the simulations having shown that the method is more sensitive to unresolved ``thin'' filaments with respect to the resolved ones independently of the contrast.  There is no apparent reason that far, unresolved, structures should not be detected by the algorithm, as long they are not so shallow to be confused with the variations of the background. We have checked that the filtering of the sample on lengths, ellipticities and filling factors does not systematically remove only the filaments with width of the order of the beam finding that there is no net effect on the width distribution of the sample.
Nevertheless, Fig.\ref{Filwidth} still show a selection effect with the larger structures identifed at farther distances. If the same filament population detected at about $\sim$1\,kpc would be shifted toward larger distances, we should detect a larger number of unresolved structures. Instead, if we compare the width distributions of the filaments separating the sample into distance bins, we found effectively a  lack of narrow structures. Part of the reason is to be attributed to the beam dilution that smooth more the variation of the density gradient for more distant objects as discussed in Sect.\ref{SecDisc} affecting the ability to detect the filaments with respect to the their surrounding emission. Furthermore, we analyzed the SPIRE maps and the column density map computed from them that we adopted in this study to search for structures with sizes of the order of the beam. We found that either for filaments and for compact sources there is a low number of objects whose width is close to the theoretical SPIRE beam. For the case of the compact sources identified by \citet{elia13} we found that the large majority of sources have a size that is $\sim$1.2$--$1.3 times the theoretical beam. A similar result is found almost everywhere in the galactic plane (see also \citet{Molinari2014}). We explain such a result measured directly on the maps as an effect introduced by the local halo in which the sources are embedded that broaden the radial profile. In fact we note that the few cases compatible with the beam are generally well isolated objects on a very low background emission. A similar effect is found also for the filaments, the structures usually embedded in dense extended enviroment have a broader profile than the few isolated filaments. If we consider as unresolved all the filaments with a width within 1.3 times the theoretical beam size we find that $\sim 20$\% of the whole sample are compatible within the errors with a structure not resolved in the radial direction. A similar percentage is found if the sample is split among filaments closer and farther than 1.5\,kpc showing that there is no significant statistical difference in the sample with the distance.
 
We computed the deconvolved widths for the resolved filaments and found a  median width equal to 0.3\,pc, a factor of $\sim$3 larger than the one identified in nearby clouds \citep{Arzoumanian2011}.

The distribution of the aspect ratio, defined as the ratio between the filament length and its deconvolved width, is presented in Fig.~\ref{hist_filaments} (panel $d$). The median value for the whole sample of filaments is 7.5, with the bulk of filaments having aspect ratios between 2 and 40. 

The filaments we have identified on the Hi-GAL maps are typically longer and have higher aspect ratios than the ones found by \cite{hacar2013} in the L1495/B213 Taurus star-forming region, which have lengths ranging between 0.2\,pc and 0.6\,pc and aspect ratio between 2 and 7. The Hi-GAL filaments are instead similar to the filaments identified by ammonia emission in more distant massive star-forming regions (i.e. \citealt{busquet2013} with lengths of 0.6--3.0\,pc and aspect ratios of 5--20). It is not unexpected to find structures of different lengths when analyzing a whole portion of the Galactic plane. However, we stress that, at least for the lower-contrast filaments, the measured lengths might be underestimated. Based on our simulations we found that some of the shorter filaments might belong to longer structures that were split into smaller portions by the adopted threshold in the filament extraction.
 
Our sample of filaments is spread over a wide range of distances, with the majority located around 1\,kpc. 
Given the almost bimodal distance distribution around 1 and 2\,kpc, we divided the sample, for which we know the distance through the association with clumps, into ``near'' distance for filaments with $d<1.5$\,kpc and ``far'' distances $d>1.5$\,kpc (see also the blue dashed line in panel ($a$) of Fig.~\ref{hist_filaments}). There are almost twice as many filaments at ``near'' distances (121), than at ``far'' distances (70). The distribution of the ``near'' filament lengths has a mean of 2.6\,pc with a standard deviation  of 2.1\,pc, while for ``far'' filaments the mean length is 6.9\,pc and the standard deviation is 8.0\,pc. ``Near'' filaments have more constrained spine lengths and are well represented by the main distribution seen in panel ({\it b}). The spine length of farther filaments has a larger spread all across the histogram (panel ({\it b})), however such an effect is mostly due to the cut on the size of the structure we have imposed in the extraction. In fact, the distribution shows a cut-off at $\sim$1.4\, pc, corresponding to our filter length of $\sim$ 2' for the median distance of 2.45\,kpc if we consider only the sample at the ``far'' distance. Our filter length implies that, depending on the distance, we are missing structures with lengths between 1.4  to 4.6\,pc, the latter being the shortest filament we would keep for the distance of 8.5\,kpc.
The filament (deconvolved) widths show a similar trend: nearby filaments widths are narrower and well confined, the distribution has a mean of 0.26\,pc and a standard deviation of 0.16\,pc, farther filaments have wider widths with a larger spread, with a mean of 0.82\,pc and standard deviation of 0.57\,pc. Again the effect of the distance justifies the two different shapes of the distribution.

\subsection{Probability density functions of column density}

\begin{figure*} 
\includegraphics[width=18cm]{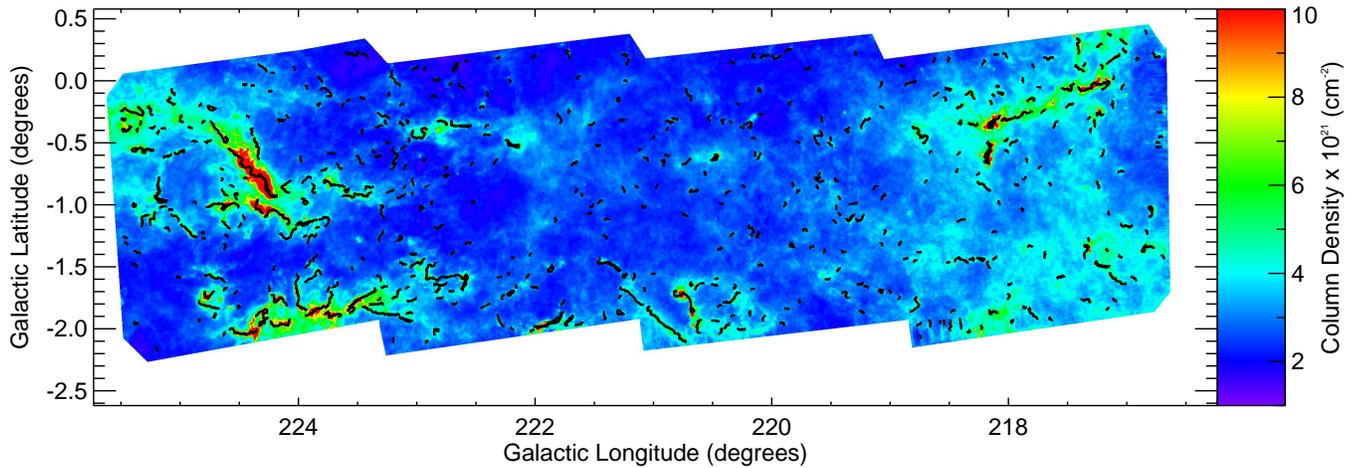}
\caption{Column density map computed from the Hi-GAL maps in the Galactic longitude range of $l$=217 to $l$=224 presented in \citet{elia13}. The thick black lines indicate the main spine of each detected filament. The detected filaments are the ones that show a variation of their contrast higher than 3 times the standard deviation computed locally on regions that are wide $11.7 \times11.7$ arcmin$^2$ in size.
}
\label{OutputDetection}
\end{figure*}

\begin{figure} 
\includegraphics[width=8cm]{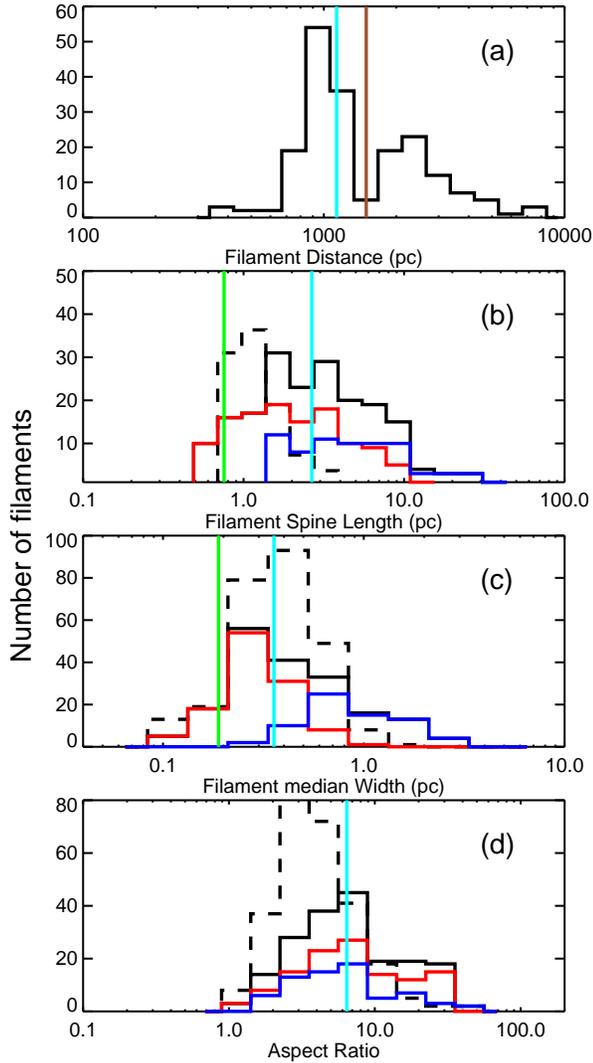} 
\caption{Distributions of filament properties: filament kinematic distance (panel $a$), filament length (panel $b$), measured width (panel $c$), aspect ratio (panel $d$) 
in the $\sim$500 filaments identified in the $l=$217--224 longitude range. The cyan vertical line depicts the median value of filaments distance (1.1\,kpc), length (2.5\,pc), aspect ratio (7.5). The red line in the panel $a$ indicates the 1.5\,kpc separation mark between the ``near'' and ``far'' sample (see text).
In panels ($b$) to ($d$) we plotted in solid lines the spine lengths, the width and the aspect ratio's for the filaments with a distance estimate while the dashed line shows the histogram of the spine lengths and aspect ratios of the filaments for which we assumed the median distance of 1.1\,kpc. Furthermore, we divided the sample depending their estimated distance in near (d $<$ 1.5\,kpc, plotted in {\it red}) and far objects (d $\ge$ 1.5\,kpc, plotted in {\it blue}). The green vertical line shows the length cutoff for filaments at 1.1\,kpc. \label{hist_filaments}}
\end{figure}

\begin{figure} 
\includegraphics[width=\columnwidth]{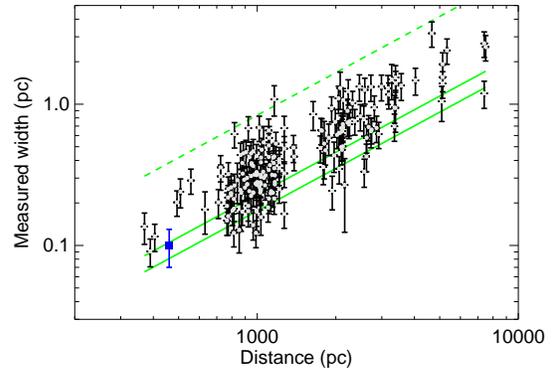} 
\caption{Measured width as a function of the distance for filaments with a reliable distance. For comparison, we also indicate with green lines the apparent size of an unresolved object with size equal to the theoretical beam (solid line on bottom), 1.3 times the theoretical beam (see the text for details - solid line on top) and five times such a value (dashed line) as a function of the distance.
The blue square represent the median size of the filaments identifed by \citet{Arzoumanian2011} in the IC5146 molecular cloud at a distance of 460\,pc.} \label{Filwidth}
\end{figure}

Fig.\,\ref{OutputDetection} strongly suggests that the filaments are denser structures with a certain morphology, sometimes embedded in a less dense molecular cloud.
Hence, before discussing the average physical properties of the identified filament sample, we discuss  the probability density functions (PDFs) of column density to quantify the difference between the filamentary structure and the more diffuse material. PDFs are a useful tool for detecting the presence of density structures, e.g. clumps and cores, in molecular clouds. 
A lognormal distribution of column densities is usually taken as proof of an isothermal medium where significant large-scale turbulent motions are taking place 
\citep{vazquez01}, while the departures from that, generally identified as power-law tails in the high column end of the distribution, are a sign that self-gravity is starting to take hold \citep{kainul09} or, equally feasible, for the presence of a non-isothermal turbulence \citep{passot1998}.

The global PDF of the $l$217--224 region has already been discussed in \citet{elia13}, here we continue the analysis dividing the region into on and off filament.
Furthermore, we separate the PDFs for pixels containing clumps from those without for both filamentary and non-filamentary regions, by flagging for each clump all the pixels within a HPBW of the clump position. Such a separation quantifies how much the clumps contribute at the high column density end of the distribution, where we expect a strong contribution from gravitationally bound structures.

\begin{figure} 
\includegraphics[height=\columnwidth,angle=90]{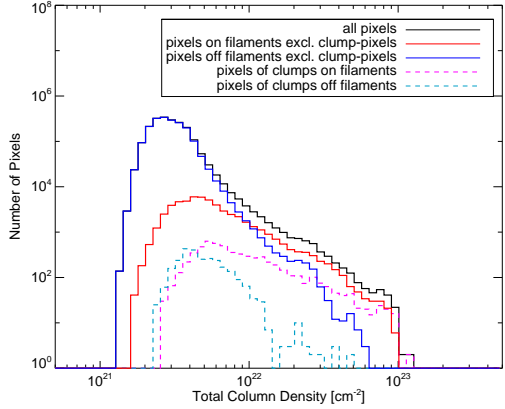}
\caption{\label{fil_histncol}  Column density map pixel distributions. Pixels associated with a filament constitute the violet histogram, Pixels not associated with filaments constitute the blue histogram. }
\end{figure}

Fig.~\ref{fil_histncol} shows that the shape of the distributions of filamentary (without clumps) and non-filamentary (with clumps) regions are clearly different. The off-filament pixels follow a lognormal distribution at low column densities with a power-law at higher column densities. The peak of the distribution is at $N_{\rm H_{2}}$$\sim$2.5$\times10^{21}$\,\cmtwo\ representing the column densities of the diffuse galactic material in the outer galaxy (see also Fig.\,\ref{OutputDetection}). 
While the filaments'pixels have a less pronounced lognormal distribution peaking at higher column densities ($N_{\rm H_{2}}$$\sim$4$\times10^{21}$\,\cmtwo), they have a much more dominant power-law tail at high column densities.  Below the limit of $N_{\rm H_{2}}$$\le$4$\times10^{21}$\,\cmtwo\ only 1\% or less of the pixels in the column density map lie in filaments. Column densities with 4$\times10^{21}$\,\cmtwo\ $\le$ $N_{\rm H_{2}}$$\le$6$\times10^{21}$\,\cmtwo\ are clearly dominated by the non-filamentary molecular cloud emission. For $N_{\rm H_{2}}$$\ge$6$\times10^{21}$\,\cmtwo\   the majority of the pixels fall in filamentary regions. 

Clumps on filaments dominate the high-column density end, as expected, but do not account for the entire power-law tail.  Hence, the dense material contained in the filament, but not in the clumps, may indicate that the filament itself is not dominantly made of isothermal material. A possible explanation is the presence of some clumps not identified in the previous analysis, while a more suggestive hypothesis would be that self-gravity is taking over not only in the clumps but also in some portions of the filament. In other words, the presence of such high density regions might indicate that filaments are part of a globally collapsing flow. Our data are not conclusive to determine if such hypothesis is correct and further investigation through spectroscopic data is needed. In fact, \citet{schneider2010} and \citet{Kirk2013}, analyzing the molecular line profiles, found hints of global collapse and accretion onto the filaments in nearby star forming regions.

The clump pixel distribution off filaments does not dominate at the high column density end ($N_{\rm H_{2}}$$\ge$\,2$\times 10^{22}$\,cm$^{-2}$), but at intermediate column densities (3$\times 10^{21}$\,cm$^{-2} \le N_{\rm H_{2}}$$\le$\,2$\times 10^{22}$\,cm$^{-2}$). It is likely that the high-column density end of the off filament distribution belongs to the dense molecular cloud surrounding the filamentary structures (see also Fig.\,\ref{SingleFilaments_pag1} and Fig.\,\ref{SingleFilaments_pag2}).

\subsection{Filaments column densities}
\label{physprop}

\begin{figure} 
\includegraphics[width=\columnwidth]{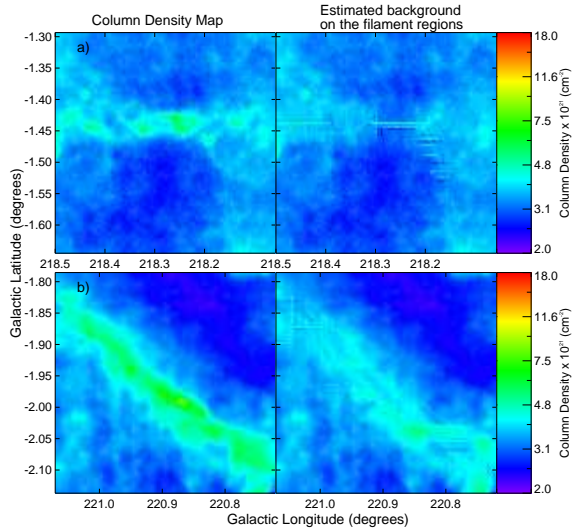}
\caption{Examples of filaments. In left panels we present the original column density map where the filaments are identified, while in right panels  we show the estimated background on the filaments. The top panels {\it a}) and {\it b}) are examples of isolated filaments, while panel {\it c}) and {\it d}) are filaments embedded in the denser enviroment 
of a molecular cloud. \label{SingleFilaments_pag1}}
\end{figure}

\begin{figure} 
\includegraphics[width=\columnwidth]{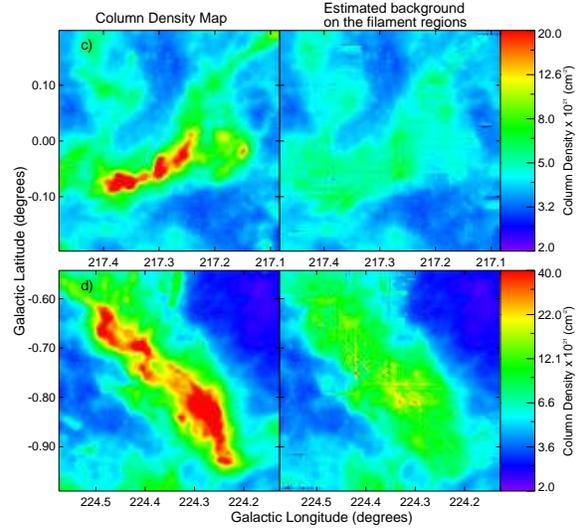}
\caption{As in figure \ref{SingleFilaments_pag2}, but for the case of filaments embedded in the denser enviroment of a molecular cloud.  \label{SingleFilaments_pag2}}
\end{figure}

\begin{figure} 
\includegraphics[width=\columnwidth]{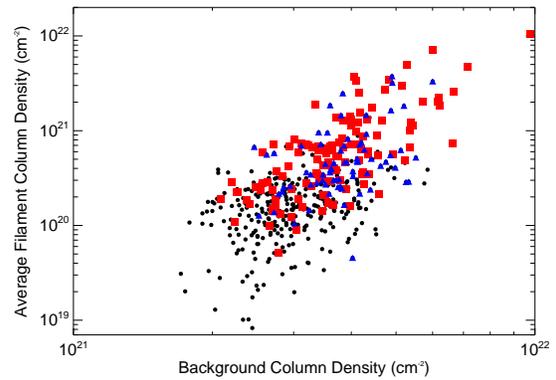}
\caption{Filament background column density against  average filament column density after subtracting the background, red squares indicate filaments with distance d$<$\,1.5\,kpc, while  blue triangles filaments d$\ge$\,1.5kpc. The black dots indicate the filaments without an associated distance arbitrary assigned at distance d$=$ 1.1\,kpc (see the text for details).
\label{fil_ncolncolback}}
\end{figure}

We have identified filaments in the {\it Herschel} data as isolated structures as well as parts of a more complex structure: the denser parts of a molecular cloud, see Fig.\,\ref{OutputDetection}.
For every pixel in each filament we defined {\it the contribution to the measured column density from the filament} as the difference between the pixel value, given by the greybody fit of the 160--500 $\mu$m fluxes, and the local estimated background given by the interpolation along the direction orthogonal to the filament spine (see also Sec.\,\ref{realestim}). 
Fig.\,\ref{SingleFilaments_pag1} and Fig.\,\ref{SingleFilaments_pag2} show some examples of filaments (in the left panels) and the estimated background (in the right panels). Panels {\it a)} and {\it b)} in Fig.\,\ref{SingleFilaments_pag1} (the latter shows the same filament of Fig.\,\ref{real_filwidth}) are isolated filaments which include the majority of the material, while in panels {\it c)} and {\it d)} in Fig.\,\ref{SingleFilaments_pag2} the identified filaments are deeply embedded in the cloud. Denser filaments are found in denser enviroment, as shown also in Fig.\,\ref{fil_ncolncolback}, suggesting a scaling relationship between the mean density of the background and the matter accumulated into the filament. 

\begin{figure} 
\includegraphics[width=\columnwidth]{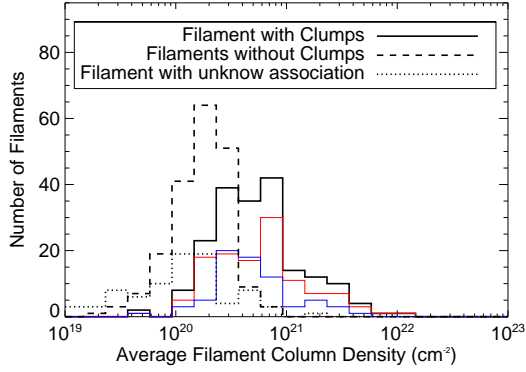}
\caption{Histogram of filament column density for filaments with and without clumps. The sample has been divided in three groups, filaments wih clumps, filaments without clumps and filaments where the association is uncertain (see text). Furthermore, the filaments with clumps have been separated into ``near'' (d$<$1.5\,kpc - plotted in red) and ``far'' (d$\ge$1.5\,kpc - plotted in blue) filaments. The black dots represent filaments without a distance estimate for which we assumed a distance d$=$ 1.5\,kpc. \label{fil_histncolclump}}
\end{figure}

We show in Fig.\,\ref{fil_histncolclump} the histogram of the {\it mean} value of the filament contribution to the column density, adopted in the following as an estimate of the average column density of the filament. 
Our filament sample covers a range of average column densities from 10$^{19}$~\cmtwo\ up to 10$^{22}$\,\cmtwo. We divide the sample into three groups: "{\it A}" filaments with at least one associated clump and therefore with a kinematic distance, "{\it B}" filaments without any association, lacking distances, and "{\it C}" filaments with unknown association (see Sec.\,\ref{secfil}), still lacking a distance determination.

Filaments with clumps are clearly denser (median value is $4.8\times10^{20}$\,\cmtwo) than those without (median equal to $1.7\times10^{20}$\,\cmtwo). 
The distribution of filaments in group "{\it A}" and "{\it B}" are very different and the probability {\it P$_{AB}$}, obtained with the Kolmogorov-Smirnoff (KS) statistic, that the populations are drawn from the same distribution is very low ({\it P$_{AB}$}$\ll$1$\times10^{-5}$). Moreover, we computed either the probability that the population with unknown association "{\it C}" could be drawn from the distribution of filaments with clumps, {\it P$_{CA}$}, and the probability that "{\it C}" could be drawn from the filaments without clumps, {\it P$_{CB}$}. We found that the population "{\it C}" is significantly different from both,  {\it P$_{CA}$}$\ll${\it P$_{CB}$}$\sim10^{-4}$, however, the large difference between the two probabilities seems to indicate that this sample is composed mostly by filaments without clumps. 
We conclude that the differences between the populations "{\it B}" and "{\it A}" are real and not due to the lack of distance determination. The filaments in population "{\it B}" did not form clumps because the column densities were too low.  

Despite the column density being a {\it distance independent} quantity, it can be influenced by beam dilution in unresolved or low filling factor sources. Given that one observes a structure with the same physical size, one expects lower column densities at farther distances when beam dilution plays a role. As in Sec.\,\ref{secfil}, we divided the sample into the ``near'' and ``far'' populations and overplotted the corresponding histograms in Fig.\,\ref{fil_histncolclump}. Indeed, we found that the average column densities for nearby filaments are higher and with a larger spread (the distribution has a mean of 9.7$\times10^{20}$\,\cmtwo\ and a standard deviation of 14.0$\times10^{20}$\,\cmtwo), while at larger distances filaments have lower column densities and with a smaller spread (the distribution has a mean of 6.8$\times10^{20}$\,\cmtwo\ and a standard deviation of 7.5$\times10^{20}$\,\cmtwo). Given the small differences between the mean values of the two samples, it appears that beam dilution has a small effect on {\it the average column density estimator}. This is especially true considering that almost all the filaments in our sample are resolved in the radial direction and have lengths several times larger than the beam. 

Our estimates of the column densities are generally lower than the ones found by \citet{Arzoumanian2011} in nearby star-forming complexes. However, we point out that they report the central column densities, which are always higher than the mean value of the overall structure. To compare these quantities we also estimated the central column densities from the pixels of the filament spine that {\it do not belong to individual compact sources}, after the subtraction of the estimated background. The maximum central column densities measured are a factor 5.8$\pm$2.4 higher than the average filament column densities. However, the maximum values for the column densities along the spine might be influenced by undetected sources and/or by local density enhancement. In the same way, the mean and the median are affected by the low column density pixels, still traced by the code, connecting different potions of the filament through regions where the material has been partially removed. If we adopt as estimator the 3$^{rd}$ quartile of the distribution of the column densities along the spine, background subtracted and not belonging to the sources, we would find that the central column densities are higher by a factor of  3.1$\pm$1.2 than the average column densities. With the adoption of these factors, we again find that our central column densities are comparable with the ones found with {\em Herschel} by \citet{Arzoumanian2011} in nearby star-forming complexes ($N_{\rm H_2}$$\sim$10$^{21}$\,\cmtwo).

Finally, we want to emphasize some caveats related to our estimation of the filament contribution to the column density.
First, in every pixel  the column density values are affected by beam dilution, which smooths the density enhancements in the central part of the filaments. 
Second, when we compute the filament contribution after subtracting an estimate of the background, consisting of diffuse emission from the Galactic plane and/or the underlying surrounding material, we are implicitly assuming that the background and filament add linearly to give the calculated column density. Strictly speaking, this is only true when both components have roughly the same temperature. However, this condition is generally not satisfied, expecially in the denser regions, which are  typically colder. The overall effect is that the assumed filamentary column density contribution in a {\it single pixel} is an underestimate of the real column density, with larger discrepancies found at higher column densities.

Both beam dilution and the uncertainity due to the single temperature approximation affect the estimate of the central column density with respect to the average column density defined at the beginning  of this section. Hence, in the following, all the derived quantities related to column density have been estimated using the average value. 

\subsection{High-mass star formation in filaments}

\citet{elia13} performed a thorough study of the star-forming content of the $l$217--224 longitude range of the Hi-GAL data. They identified the compact sources (clumps), and classified them as protostellar, prestellar, or unbound clumps. Protostellar objects are objects which contain 22 and/or 70 \um\ emission indicating the presence of young stellar objects (YSOs). The remaining starless objects, which do not contain a YSO (no 22 or 70 \um\ emission), can be divided into prestellar objects, which are gravitationally bound objects evolutionarily younger  than protostellar ones, and unbound starless objects, by comparing their mass, $M$,  with their Bonnor-Ebert mass, $M_{BE}$ (see \citet{elia13} for a detailed discussion). Here we adopted the criteria of $M > M_{BE}$ to separate the prestellar objects from the unbound starless ones.

We correlate the above clump classification with our filament sample to understand the impact of filamentary structure on the star formation activity.
The overall result shown by Fig.\,\ref{fil_histncolclump} is that star formation is  found preferably on the densest filaments, with higher average column densities (distribution peak is at 0.5--1$\times10^{21}$\,\cmtwo) than those without star formation, whose average column densities peak at $2\times10^{20}$\,\cmtwo. The differentation starts at 4--6$\times10^{20}$\,\cmtwo\ (see Fig.~\ref{fil_histncolclump}). Moreover we found that not all filaments have clumps. If we exclude from the sample the filaments for which the clump detection might have been biased due to the NANTEN observation coverage and sensitivity, we found that 50\% of the detected filaments do not have clumps. These filaments without clumps are in a very early stage of evolution or, alternatively, they might be transient structures, only confined by the external pressure.

Not all the remaining filaments show signs of ongoing star formation, in fact filaments with only unbound starless clumps (4\% of the sample) do not contribute to it. 
 
\citet{Polychroni2013} investigated the fraction of  sources {\it on} and {\it off} filament in the L~1641 clouds in the Orion A complex and found that 67\% of the prestellar and protostellar sources are located on a filament. In our case, that includes several molecular clouds, we find a similar fraction with the majority, 74\%, of the clumps reported by \citet{elia13} falling within our filament sample.  However, there are still a significant number of clumps detected off filaments.

\begin{figure} 
\includegraphics[height=\columnwidth,angle=90]{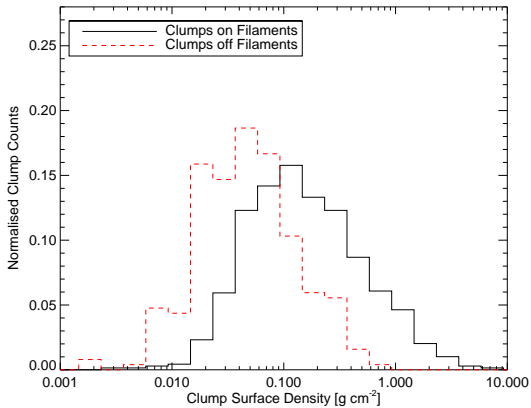}
\caption{Clump surface density distribution for clumps located within filaments and clumps not located within filaments. \label{fil_histsurf}}
\end{figure}

We computed the clump surface densities from the masses and radii reported in \citet{elia13}  and compared clumps located on filaments, $\Sigma^{on}$, and these not on filaments, $\Sigma^{off}$ (see Fig.~\ref{fil_histsurf}). The distribution of $\Sigma^{on}$ peaks around 0.1\,g~\cmtwo\ and reaches values  up to 10\,g~\cmtwo, while $\Sigma^{off}$ peaks below 0.1\,g~\cmtwo\  and it reaches a maximum value of $\sim$0.8\,g~\cmtwo. The shape of the two distributions are statistically different with a more evident tail toward higher surface densities for sources on filaments. The different shapes might be explained by the larger uncertainities of $\Sigma^{on}$ due to the difficulties in decoupling of the compact source contributions from the underlying structure. However, we estimate that such uncertainities are up to $\sim$30\%, while to match the two distributions is would be needed that $\Sigma^{on}$ is overestimated by a factor of $\sim$3. Therefore, it is very likely that the sources on the filaments have larger surface densities than the ones outside. More important, we observe that all the sources lying outside of the filamentary regions have surface densities smaller than $\Sigma \sim$ 1\,g~\cmtwo . Such value is advocated in theory of the star formation through turbulent accretion as the threshold limit below which the {\em massive} stars cannot form due to fragmentation.
Recent observations indicate that such a threshold limit should be revised to a lower value of $\sim$0.2\,g~\cmtwo \citep{Butler2012}. Even with this revised limit, our results suggest that it is favourable to form massive stars in the filamentary regions.

Similar conclusions were reached by \citet{Polychroni2013} from the analysis of the clump mass functions (CMFs) for {\it on} and {\it off} filament sources. They found, indeed, that the CMF of {\it on}-filament sources peaks at higher masses ($\sim$ 4\,\msun) than the {\it off}-filament ones ($\sim$ 0.8\,\msun), suggesting that the discrepancy is caused by the larger reservoir of material available locally on filaments in respect to the isolated clumps.

\subsection{Stability of filaments}

Given that filaments can be roughly approximated as cylinders, theory shows that such structures have a maximum linear density, or mass per unit length \mline, above which the system would not be in equilibrium against its self-gravity. For the simplified case of a cylinder infinitely extended in the {\it z}-direction with support given only by thermal pressure, the critical mass per unit length, \mlcrit, is only a function of temperature \citep{ostriker64,larson85}. It becomes a more complicated function when other effects like turbulence and/or magnetic fields are taken into account  \citep{Fiege2000}, in this case the \mlcrit\ will increase by a small factor. Structures with \mline above the \mlcrit will start collapsing along the radial direction.
The critical mass per unit length scales linearly with the temperature and its value is around \mlcrit$\sim$16~\msunpc\  for the typical temperature in molecular clouds of $T$\,$\sim$10~K, \citep[see for example][]{Andre2010}. 
We estimated the mass per unit length for each detected filament from the average column density in the RoI, given as the sum of the contribution from the filament (see Sec.\,\ref{physprop}) divided by the area of the RoI, times the mean width. In the simple case of a straight filament, aligned on the plane of the sky, with no density variations along the spine and constant radial profile along the structure, the estimator defined above equals the integration along the radial profile divided by the length of the structure, i.e. the mass per unit length. The unknown inclination of the structure affects our estimate of \mline. However, while on the one hand the area of the filament projected on the plane of the sky is reduced due to projection effects, on the other the measured column density will increase by $\sim$60\% for the same reason. The two effects partially balance each other.

For the more complex filaments detected in this field there might be small discrepancies. We stress that our definition entails an average (global) \mline\ for the whole filamentary structure. In other words, we are assuming  {\it the total mass measured in the entire structure was initially uniformly distributed along the filament when it formed}. Therefore, even if we determine low values of \mline\, we cannot  exclude that locally, in a small portion of the filament, the density is enough to become gravitationally unstable. 

\begin{figure} 
\includegraphics[width=\columnwidth]{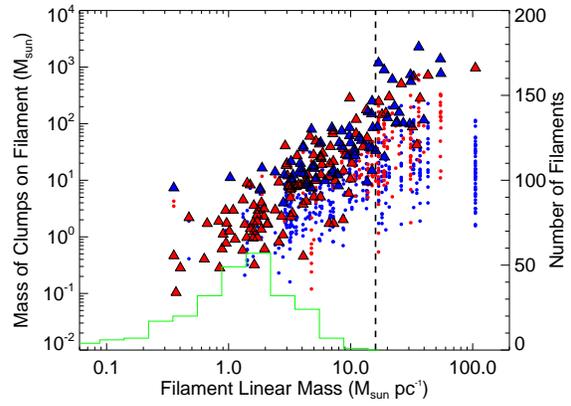}
\caption{Mass of the clumps inside a filament as function of the hosting filament mass per unit length. Small dots are individual clump masses divided in red for near filaments (distance d$<$1.5\,kpc) and blue for far ones (distance d$\ge$1.5\,kpc).  Triangles depict the total mass in clumps on that filament. For  comparison, we also plot the filaments without clumps (green histogram) to show their distribution in mass per unit length. The vertical dashed line marks the critical mass per unit length \mlcrit$\sim$16~\msunpc\ for $T$\,$\sim$10~K. \label{fil_lmclumpmass}}
\end{figure}

Figure~\ref{fil_lmclumpmass} shows the mass of the clumps identified in the filaments with respect to \mline. We have excluded from our analysis the filaments with starless clumps that have no contribution in terms of star formation. We additionally show the distribution of \mline\ for these filaments without clumps (green line): all filaments without clumps have \mline$<$10~\msunpc, \ie\ smaller than \mlcrit$\sim$16~\msunpc. Filaments with clumps span a much wider range of values, with \mline\ up to 100~\msunpc. Many filaments found in the $l$217--224 field have \mline$<$16~\msunpc, and hence they are subcritical (\mline$<$\mlcrit), even though they contain clumps.  We stress that the value of 16~\msunpc should not be taken as a strict limit since real filaments {\it a)} might not be correctly described by isothermal model and {\it b)} have a finite extension along the {\it z-}direction. Hence, even in the case of finite slightly ``subcritical'' filaments it is expected that both the external pressure and the gravity play a role. 
 
The distribution of \mline\ for filaments with clumps (see Fig.\ref{fil_lmclumpmass}) indicates that many filaments did not have initially supercritical \mline\ to begin the clump collapse and star formation. If we assume that \mline\ remained unchanged during the onset of star formation, then this implies that gravity alone was insufficient to cause a global collapse into clumps. Moreover, the ability to form clumps with lower or higher masses is inherent to the initial \mline\ of the filament:  more massive clumps form in the more critical (and supercritical) filaments.  Since this last result might be biased by including the contribution of the clumps to the average column density,  we estimated the \mline\ excluding the clump mass contribution to the identified filament. 
This estimator represents the stability of  the remaining material in the filament against gravitational collapse and for a filament where the star formation process is complete it can be lower than \mlcrit. The trend shown in Fig.\,\ref{fil_lmclumpmass} does not change with removing the clumps from the \mline\ calculation, indicating that the relation between \mline\ and the clump mass is real.

Finding filaments hosting clumps with an average \mline\ lower than the critical value is not completely surprising. Our filaments are comparable  
to infrared dark clouds (IRDCs) and high-extinction clouds that often display a filamentary morphology and are generally found to have distances of 2--4\,kpc \citep[see for example][]{rathborne2006, rygl2010}. Recently, \citet{Hernandez2011} and \citet{hernandez2012} investigated the dynamical state of two IRDCs and found \mline$\simeq$0.2--0.5\mlcrit. While their analysis is based on molecular line data, taking into account the (stabilizing) non-thermal contribution to the \mlcrit, they find clear signs of star formation activity, through the presence of 8~\um\ and/or 24~\um\ point sources, in gravitationally stable structures.  
More generally molecular clouds, and also filamentary molecular clouds, are found overall to be gravitationally unbound when their masses are compared to the virial mass despite the fact they contain gravitationally bound clumps and star formation \citep{rygl2010,Hernandez2011}. The general idea is that these large scale structures are not far from virial equilibrium and that external pressure or flows could have initiated the star formation activity \citep{Tan2000}. Therefore, while the external pressure is confining the larger structures, the smaller scales (found locally) have to be supercritical to show hints of star formation activity \citep{Andre2010}.
The sweeping up of interstellar material and its accumulation in large scale filamentary structures through converging flows \citep{Heitch2008,Vazquez2011} is compatible with such results, forming the large structure and the inital local overdensities   at the same time. Our results indicate that these processes have to act quickly,  on timescales shorter than the filament and clump free-fall time scales. If that were not the case, supercritical filaments with \mline $>$ \mlcrit\ would have a larger number of clumps, since further clumps forms as result of the filament contraction and fragmentation. We do not find such an evidence in our sample.
Even though it is difficult to confirm the converging flow scenario without kinematical information or shock tracing molecules, such as narrow SiO emission, \citep{JimSerra2010}, our data encourage the further investigation of these converging flows. Furthermore, we expect that all the supercritical filaments are characterized by a state of global gravitational collapse, such as the DR\,21 filament \citep{schneider2010}, where molecular line observations strongly suggest the convergence of large scale flows as the cause of its formation \citep{schneider2010, csengeri2011}. 

\subsection{The nature of filaments}

In the previous section we concentrated on average global quantities measured on the whole filamentary structure.
However, as explained above, the evidence of subcritical filaments with hints of star formation requires the presence of local instabilities inside the filament.
Therefore we  focus the following analysis on filament branches (see Sec.\,\ref{performance} for the definition) which give more local information than the global averages described so far.
 
\begin{figure} 
\includegraphics[width=\columnwidth]{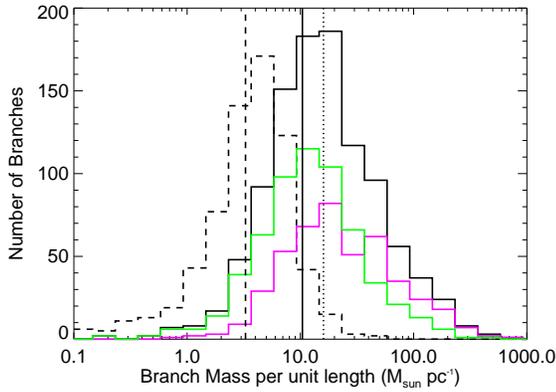}
\caption{Histograms of branch mass per unit lengths for the branches belonging to filament with clumps (solid black line) and filament without clumps (dashed black line) with their median represented as vertical lines. The branches belonging to filament with clumps are further splitted into branches hosting clumps themselves (magenta line) and branches without clumps in their local surroundng (green line). The dotted line indicate the \mlcrit value of 16\,\msunpc. \label{bra_histlm}}
\end{figure}

Figure~\ref{bra_histlm} shows the histograms of the branch \mline\, after separating the branches into filaments hosting clumps from the branches in filaments without clumps. The branches belonging to filaments with clumps have higher linear densities (median \mline\ $\sim$10 \msunpc) than the ones without (median \mline\ $\sim$3.3 \msunpc). 
Furthermore, we split the branches within filaments with clumps on the basis of their local association with clumps (in green) or not (in magenta). We found that the branches without clumps, but belonging to filaments with clumps, have still larger linear densities (with a median \mline\ $\sim$10 \msunpc) than the branches without any clumps in their surroundings. It is unlikely that this difference is due to the distance association, even if the \mline\ depends linearly on the distance, since we found that a  similar relationship exists for the branch average  column densities. 
The branches with clumps, instead, are denser with a median \mline\ $\sim$ 17\msunpc. They dominate the distribution for \mline$\ge$16\,\msunpc\ (the equilibrium limit against fragmentation for an isothermal cylinder at 10~K,  dotted line), despite the fact there are still a few branches without clump closer to \mlcrit.

\begin{figure} 
\includegraphics[width=9cm]{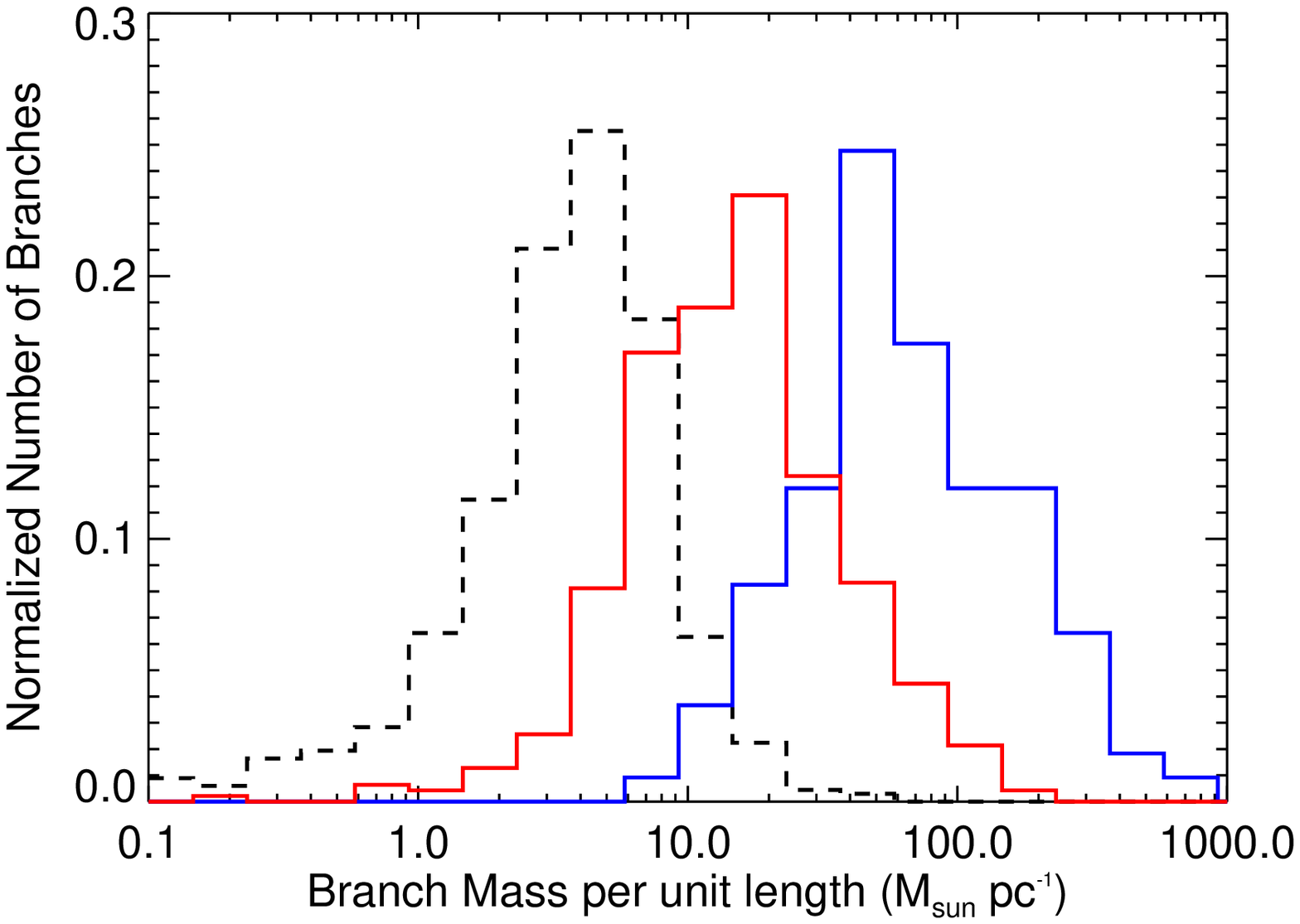}
\includegraphics[width=9cm]{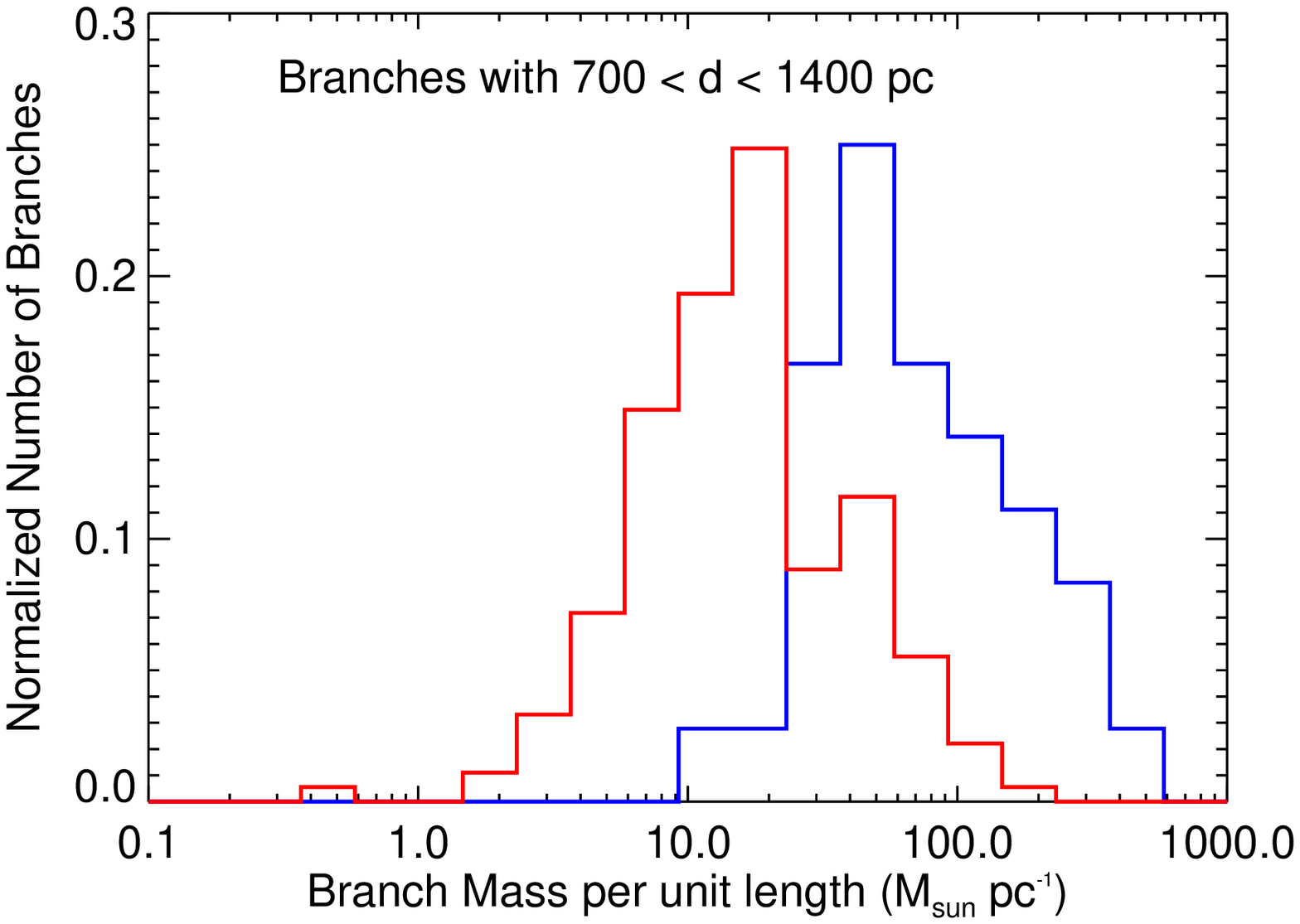}
\caption{{\it Top panel:} Histogram of masses for unit length of all the branches with protostellar clumps (blue solid line), with prestellar clumps (red solid line), and branches without clumps (black dot dashed line). {\it Bottom panel:} Distribution of masses for unit length of the branches with associated distances between 700 and 1400 pc with protostellar clumps (blue solid line) and prestellar  clumps (red solid line). \label{bra_lmevo}}
\end{figure}

We have further refined the division in Fig.~\ref{bra_histlm} by specifying the branches that contain protostellar objects, prestellar objects, and branches without clumps in the top panel of Fig.\,\ref{bra_lmevo}. For the latter we took only the branches where we can determine a distance through filament association to avoid any bias that might result from the assumed distance.  The classification of a branch is based on the most evolved object found within, hence, branches that contain both prestellar and protostellar objects have been considered as branches with protostellar objects. With this definition, the total number of  branches with an associated clump divides into  20\% classified as protostellar and 80\%  classified as prestellar.  
We found that branches with protostellar clumps have the highest \mline\, with a median value of $\sim$60\,\msunpc, well above the critical mass per unit length of 16\,\msunpc . Branches with prestellar clumps have a median \mline\ $\sim$15\msunpc .   We further checked if our results are affected by systematic effects due to the distance since we are integrating on larger volumes for more distant filaments. In the bottom panel of Fig.~\ref{bra_lmevo} we select only the branches of the filaments that falls in the range between 700 and 1400 pc. We chose such an interval to have a statistically significant number of objects. The difference in \mline between branches with protostellar clumps and the ones with prestellar clumps is mantained. In such a distance range we found 180 branches (84\% of the total) with prestellar clumps and 36 (16\% of the total) with protostellar. We tested different distance ranges and found that the branches with prestellar clumps always have a distribution with a median between 10 and 14\,\msunpc, while the median of the distribution of branches with protostellar clumps varies between $\sim$40 and 70\,\msunpc. The number fraction of branches classified as prestellar is always 4-5 times larger than the one classified protostellar. We conclude that distance selection effects are not affecting our result.

The analysis of the local \mline\ confirms that almost all the branches hosting protostellar clumps are locally unstable against gravity despite the possibility that the overall filamentary structure being potentially  subcritical. 
The branches belonging to filaments with clumps have a higher \mline\  with respect to the branches in filaments without clumps, with values closer to virial equilibrium. 
Thus, clump formation is somehow linked to the properties of the large scale filament and, although this result might be affected by undetected clumps on the filament, the filaments hosting clumps are locally different than the ones without.
Furthermore, we interpret the strong differentiation in \mline\ for branches with prestellar and protostellar objects shown in Fig.\,\ref{bra_lmevo} in terms of an evolutionary scenario, in which the protostellar branches are intrinsically more evolved than prestellar branches. This result indicates that  \mline\ might be {\it not} constant during the onset of the star formation as also suggested by \citet{Heitsch2009}. In such a scenario the branches and their filaments increase their linear density with time by contracting and/or accreting of material. The idea of mass accumulation with time is consistent with observations of velocity shifts  along filaments \citep{Peretto2013,Kirk2013}. Moreover, Herschel observations of filaments in the Taurus star forming region revealed  substructures (``striations'') connected to the filament along perpendicular directions with respect to the filament axis \citep{Palmeirim2013}. These authors suggested that the presence of such structures are a hint that accumulation of mass is going on.

Our data indicate that the branches would have to increase their linear density by $\sim$ 45\msunpc\ (given by the difference between the medians of the two distribution) on the timescale of prestellar evolution until the first protostars forms. Unfortunately prestellar lifetimes are very uncertain,  \citet{Motte2007} suggested that it lasts for  $10^{3}$--$10^{4}$yr depending on the mass of the clumps and with the assumption of a constant accretion rate. However, observations of nearby clouds indicate that there are a similar number prestellar and protostellar cores suggesting they have a similar lifetime of the order of 4.5$\times10^{5}$yr \citep{WardThompson2007,Enoch2008}. The same lifetime of $\sim$10$^{5}$yr is estimated from numerical hydrodynamical simulation \citep{GalvanMadrid2007,Gong2011}.  
Hence, our results require an accretion rate of $\sim10^{-4}$\msunpc\ yr$^{-1}$ to match the measured higher linear densities in the filamentary region with protostellar clumps. 
Such accretion rates are possible and they are lower than that estimated by \citet{Kirk2013}, both for accretion {\it along} the filament or for accretion {\it from} the enviroment. Recently \citet{Gomez2014} analyzed filaments that form in simulations where colliding flows are responsable for the inital cloud formation. They found that filaments accrete from the enviroment and simultaneously accrete onto the clumps within them. They determine linear densities increments up to $\sim 3 \times 10^{-5}$\msunpc\ yr$^{-1}$, a few times lower than the value we estimate. However, we point out they are analyzing structures that are $\sim$15\,pc long, while our results are estimated from the measure of the local branches that are smaller portion of the filaments.

\begin{figure} 
\centering
\includegraphics[width=\columnwidth]{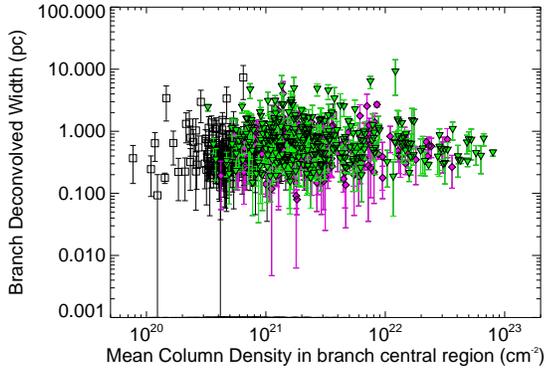} 
\caption{Mean column density measured along the central region of the branch excluding the overdensities due to the clumps versus the deconvolved width for branches belonging to filaments without clumps (black empty squares), branches without clumps but belonging to filaments with clumps (green triangles) and branches with clumps (blue diamonds). \label{bra_lmfwhm}}
\end{figure}

The observed increase in linear density in our sample could be explained by a contraction of the filaments by gravity. In such a case the shrinking of the filaments would be evident from the structures with higher central column density having smaller widths. Fig.\,\ref{bra_lmfwhm} shows the branch deconvolved width, $W$, as a function of the average column density along the branch spine excluding the source contributions, $\overline{N^{c}_{\rm H_2}}$. No hints of correlation between the width and the $\overline{N^{c}_{\rm H_2}}$ is found, regardless of whether we analyze branches with or without clumps. We obtain similar results if we select only the branches in a small range of distances to minimize the effect that we might be missing narrow, unresolved, or wide, but faint, structures. 
Hence, we rule out that contraction of the filament is responsible for the increase in linear density. A similar result was found by \citet{Arzoumanian2011}  for the filament sizes in nearby star forming regions and these authors adopted it as evidence in favour of the large-scale turbulence scenario for the {\it formation} of the filaments. In fact, if filaments were formed as a result of gravitational collapse and fragmentation of an {\it isolated sheet} \citep{Inutsuka1992}, then $W$ should be anticorrelated with $\overline{N^{c}_{\rm H_2}}$ and should be equal to the thermal Jeans length $\lambda_{J}\,=\,c^{2}_{s}/(G\mu m_{H}\overline{N^{c}_{\rm H_2}})$, inconsistent with what is found. In contrast to \citet{Arzoumanian2011} whose average width is $\sim$0.1\,pc, we measure an average filament width of $\sim$0.5\,pc in accordance with the sizes measured for some filamentary IRDCs \citep{Jackson2010}.   
Such a larger average width has dynamical implications for these structures. If we assume that the filament width is roughly equal to the effective Jean length $\lambda^{eff}_{J}$\,=\,$\sigma_{tot}^{2}/(G\mu m_{H}\overline{N^{c}_{\rm H_2}})$, where $\sigma_{tot}$ takes into account both thermal and non thermal contributions, we do expect that the supercritical filaments needs a larger non-thermal contribution to hold their sizes in comparison to the smaller structures identified in nearby star forming region \citep{Arzoumanian2013}.  An additional contribution to the velocity dispersion with respect to the one initially given by the large-scale turbulence in the interstellar medium was already suggested by \citet{Arzoumanian2013}. In our case, either these structures are formed through the turbulent scenario with larger non-thermal support or they require a larger contribution from the accretion.   
We want to stress that, since \mline$\sim$$\overline{N^{c}_{\rm H_2}}\times W$, the larger average widths imply a lower value of central column density to reach the critical value of \mlcrit$\approx$16\msunpc. Thus, we estimate that the critical central column density, excluding any non-thermal support, is $\sim$1.8$\times$10$^{21}$\,\cmtwo, implying that accretion has to be present in all regions of the filaments with $\overline{N^{c}_{\rm H_2}}$ above such value. However, such results have to be confirmed through direct measurements of  $\sigma_{tot}$ inside these regions, that is beyond the reach of the current NANTEN data. 

\begin{figure} 
\includegraphics[width=\columnwidth]{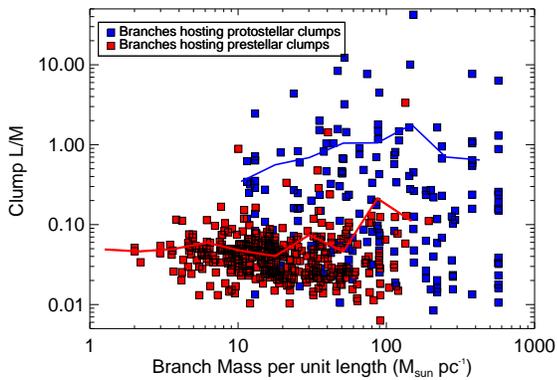}
\caption{Branch mass per unit length versus clump $L/M$ ratio\label{bra_lmratio}, with a separation between the branches hosting only prestellar clumps and the one with at least one protostellar clump. The lines represent the mean of the $L/M$ distribution in logarithmic bin of size 0.2\, dex. }
\end{figure}

Finally, in Fig.~\ref{bra_lmratio} we plot the $L/M$ ratio of the clumps versus the mass per unit length, separating the protostellar clumps (blue squares) from prestellar clumps (red squares) to provide an evolutionary marker for the objects forming in the branches. 
The $L/M$ ratio is usually correlated with the evolutionary state of a forming star (\citealt{molinari2008}). In this star-formation model, two phases are considered: 1) the mass accretion phase, in which the stellar objects accretes from its massive envelope and increases its luminosity. This phase ends when the stellar object arrives at the zero age main sequence and becomes a star; 2) the envelope dispersion phase, in which  the envelope mass decreases while the luminosity stays constant. Assuming that all the objects belong to the same initial mass function, and that each clump forms a single object, one can use the $L/M$ ratio to distinguish between more and less evolved objects.

The prestellar clumps have a constant mean $L/M$ ratio of $\sim$0.06 covering a range of local\ \mline\ between 1--100 \msunpc . The $L/M$ dispersion increases for \mline$>$ 10\,\msunpc\ due to the presence of prestellar clumps with $L/M$$>$ 0.1. Almost no prestellar clumps are found in branches with \mline\ $\ge$ 100 \msunpc.
The protostellar clumps have a higher $L/M$ ratio with median value $\sim$1 and present a larger scatter than the prestellar clumps. All the branches hosting the protostellar clumps have \mline$\ge$\,10\,\msunpc. 

If evolution is the only mechanism in action, we do expect to find higher $L/M$ ratios in branches with higher \mline . A tentative increase of $L/M$ can be drawn at least for supercritical branches that appear to be more evolved than the subcritical ones. However, the large dispersion in the $L/M$ suggests that the scenario is not complete. In particular, we cannot exclude the possibility that the branches have a different star formation rate and, therefore, the clumps  would have different values of $L/M$ despite having a similar evolution history. 
 
We found very few branches that are highly supercritical (\mline$\ge$100\,\msunpc). Surprisingly, they host very few prestellar clumps, while we would expect a larger number due to the filament fragmentation, since they evolve on faster timescales than the filament \citep{toala2012}. It is possible that the already formed clumps have a strong effect on the filamentary structure through outflows and other feedback mechanisms while the filaments themselves are still accreting material from the surroundings.  Such feedback changes the local properties of the filament and prevents the further formation of clumps. The end result is that the filament would be dissipated rather quickly. This would explain the larger number of branches with prestellar objects versus the ones with protostellars. 
Therefore, filamentary structures are rather short-lived entities.

\section{Conclusions}
\label{SecCon}

In this paper we have described a method to identify filaments of variable intensity from 2D in an automatic and unbiased way, taking into account the extended nature of these structures.
The method has been optimized to work for the typical properties of filaments observed by the {\em Herschel}, with filaments overlapping, a strong and variable background and hosting  compact sources.
With the help of simulations, we have shown the strengths of our approach not only as a way to detect, but as a tool to determine the morphological and physical properties of the filaments. Lengths and widths are typically recovered within 20\% of the expected value. Larger discrepancies are found in the case of fainter, less contrasted, structures.
We stress the need for a good estimate of the background for accurate filament mass measurements and we have shown that our approach allows us to decouple the filament contribution from the background with errors estimated around $\sim$15-20\%  (dispersion of the residuals) in the case of features with moderate surface-brightness contrast.

We applied our method to the column density map calculated from {\em Herschel} observations of the Outer Galaxy in the Galactic longitude range of $l=216.5^{\rm o}$ to $l=225.5^{\rm o}$ ($l217$--224) to measure the filament properties and attempt to determine their role in the star formation process. We found that filaments are found at various distances between 0.5 to 7.5\, kpc. 
They can be identified at various spatial scales, from lengths as short as $\sim$0.5\,pc up to 30\,pc and widths between 0.1\,pc to 3\,pc, most of which are typically resolved in the {\it Herschel} observations.The measured aspect ratios range between 3 to 30. Distances appear not to have any selection effect if not that the shorter filaments  are undetected at the farther distances.

The column density PDF indicates that almost all the dense material with $N_{\rm H_{2}}$$\ge$6$\times10^{21}$\,\cmtwo\,  is arranged into filamentary structure.
However, not all high density material is associated with clumps {\it hosted} by the filamentary structure; a significative fraction are local density enhancements on the filaments hinting of a state of global collapse. 

The majority of the clumps (74\%) identified in previous studies are located within the borders of our filament sample. Nevertheless we still found star formation going on {\it outside} the filaments. It is unlikely that these objects form in the filaments and are successively dispersed. However, we find a significative difference between the surface density of clumps on the filaments versus the one outside, with the clumps on the filaments showing higher surface densities.Furthermore, we observe that the majority of the clumps outside the filamentary regions have surface densities below the value necessary for high-mass star formation \citep{krum08,Butler2012}, hence it is very likely that they finally fragment into cluster of low-mass stars. This seems not to be the case for the clumps on the filaments where the higher surface densities overcome the assumed threshold limit. It is worth noting that so far observation of regions with high mass clumps have generally highlighted a larger filamentary structure in which they are embedded \citep{schneider2010, Hill2011, Peretto2013,Polychroni2013,NguyenLuong2011}.  
Hence, even if our data are not conclusive, the filamentary shape seems to be an important vehicle to channel enough material into small regions and to allow the formation of high-mass stars.
Observations with higher spatial resolution toward the clumps outside our filament sample would determine if massive star formation happens also without filaments and, if that is the case, the relative number of those sources with respect to the number of massive clumps inside filaments.

On the other hand, a significant number of filaments do not host any clumps. We estimated the mass per unit length, \mline , for all the filaments in our sample and found that the structures without clumps all have a \mline$\le$16\,\msunpc, the critical value  for a filament sustained by the thermal pressure exterted by material with $T\sim$10\,K. Hence, all these filaments are transient structures that are kept together by external pressure from the interstellar medium. 
The filaments hosting clumps, instead, span a large range of \mline , between 1 to 100\,\msunpc . Such a result is puzzling, since if the clumps are a direct result of filament collapse and fragmentation due to gravity, we would not expect to find clumps on any filament with \mline$\le$16\,\msunpc . Any non-thermal contribution, not accounted for by our analysis,  would increase the value of the \mlcrit  and so the number of filaments with no clumps. Such results agree with what is found for a few IRDCs  in the inner Galaxy. Moreover, we found that the supercritical filaments, \mline$\sim$80-90\msunpc\ have more material aggregated in clumps, hosting also the most massive ones.

Thus, we suggest a possible scenario for fast formation of the filaments, where these structures and the initial seed for the clumps are formed at the same time. In such scenario the global structure can be in equilibrium (or close to it), and the clump formation starts on local scales, induced by processes like flows or external pressures that locally enhance the linear density.
We studied the local scales in the filament, by analyzing the branches in which the filament can be divided. We confirmed that the branch \mline\ values are typically higher than those of the whole structure. This  result is also in agreement with the protostellar objects only found on the supercritical filaments. Structures are created with a range of masses depending on the amount of surrounding material. In the most dense ones a mini-``starburst'' process starts with a higher star formation rate that can form protostellar objects very quickly. This scenario effectively decouples the clump formation from the filament evolution (at least in their early stages). 

However, when we compare \mline\ of the local substructures we found a statistically significant difference between the filamentary subregions hosting prestellar clumps versus the one with protostellar clumps. In particular, we found higher values of linear density in the subregions with protostellar clumps with respect to the ones with only prestellar clumps. While these results are still consistent with the  fast formation scenario, it suggests that the differentation might be set by the evolution of the structure.
Filaments, after their formation, increase their \mline\ and progress to form prestellar clumps that evolve into protostars on timescales faster than the filament evolution. It is unlikely that the enhancement of \mline is due to the shrinking of the filament due to self-gravity, but our results play in favour of an accretion of material from the surrounding (within the filament itself or from the enviroment). Our data requires moderate increase of linear density with time $\sim$10$^{-4}$\,\msunpc\,yr$^{-1}$,  a rate compatible with the one measured for filaments in nearby molecular clouds.

Following \citet{Arzoumanian2013} we expect that all the filaments with widths larger than the Jeans length should be in a state of global collapse due to their gravity, since the internal thermal pressure alone is not able to sustain the structure. For the mean width of our filament sample of  0.5\,pc, we expect that all the structure with a central column density higher than $\sim$1.8$\times$10$^{21}$\,\cmtwo would be in dynamical collapse. However, since we do not measure any shrinking of the filament, hence we expect an increase of non-thermal support for those more condensed filaments. If such a result is really hinting at filaments in a state of global collapse, we do expect a larger number of those with respect to the one identified for the studies of nearby clouds  \citet{Arzoumanian2013}.  This result requires additional confirmation for future molecular spectroscopic observations.

\begin{acknowledgements}
The authors are grateful to the anonymous referee for the useful comments which improved the presentation of the work. E.S. acknowledge support from the NASA Astrophysics Data Analysis Program grant NNX12AE18G.
K.L.J.R. and G.B. are supported by an Italian Space Agency (ASI) fellowship under contract number I/005/11/0. 
DP is funded through the Operational Program ``Education and Lifelong Learning'' that is co-financed by the European Union (European Social Fund) and Greek national funds.
S.C.O.G. acknowledges support from the Deutsche Forschungsgemeinschaft via SFB 881 ``The Milk Way System'' (sub-projects B1 and B2).
The authors are also grateful to M. Pereira-Santaella and N. Marchili for helpful suggestions.  \\
E.S. is very grateful to Antonia Pierni and wishes to thank her for all the moments they spent together while he was working on this research.
       
\end{acknowledgements}

%
%
%

\clearpage

\clearpage

\clearpage

\clearpage

\end{document}